\documentclass[12pt,epsf]{article}

 \usepackage{epsfig}
 \usepackage{times}
 
\begin{document} 
     \setcounter{totalnumber} {5}
     \begin{titlepage}
 \bigskip
 \bigskip\bigskip\bigskip\bigskip
 \centerline{\LARGE Through the Looking Glass:}
 \centerline{\Large AdS-FT with time dependent boundary conditions and black
 hole formation}
 \bigskip
         \bigskip\bigskip
                 \bigskip\bigskip

 \centerline{\large Keith Copsey}
      \bigskip\bigskip
  \centerline { \em Department of Physics, UCSB, Santa Barbara, CA 93106}
    \centerline {keith@physics.ucsb.edu}

 \begin{abstract}  I solve for the behavior of scalars in Lorentzian AdS
 with time dependent boundary conditions, focusing in particular on the 
 dilaton.  This corresponds, via the AdS-CFT correspondence, to 
 considering a gauge theory with a time dependent coupling.  Changes 
 which keep the gauge coupling nonzero result in finite but physically 
 interesting states in the bulk, including black holes, while sending 
 the gauge coupling to zero appears to produce a cosmological singularity 
 in the bulk. 
 \end{abstract}

 \end{titlepage}

 \baselineskip=18pt

 \setcounter{equation} {0}
 \section{Introduction}

 \indent  Most work on AdS assumes one puts constant boundary 
 conditions on the bulk fields at spatial infinity.  This corresponds 
 to putting reflective boundary conditions (i.e. a mirror) at infinity so no energy 
 goes through the boundary.  In the context of the AdS-CFT correspondence 
 non-constant boundary conditions for the dilaton corresponds to a 
 time dependent gauge coupling in the boundary field theory.   This 
 suggests non-constant boundary conditions for the dilaton may be 
 physically interesting.  On the field theory side non-constant 
 boundary conditions imply we no longer have 
 time translation symmetry, energy conservation, conformal symmetry 
 (the gauge coupling sets a scale), or supersymmetry ($\{ \mathrm{Q}, \bar{\mathrm{Q}} \} = 
 \mathrm{P} $ and since we have no time translation symmetry we have 
 no superalgebra either).   This field theory requires investigation, 
 although very slow or fast changes in the gauge coupling would appear to be well described in the gauge theory by quasi-static and sudden 
 approximations respectively.
 
 \indent With these facts in mind I set about investigating the solution 
 in the bulk for a massless field with time dependent boundary 
 conditions.   At first one might expect even changing the gauge 
 coupling slightly would produce a rather 
 severe response in the bulk.   AdS has an infinite volume and area near the 
 boundary and hence it seems likely even a finite change in the boundary conditions 
 would send infinite energy into the bulk.  Since the field is 
 massless and AdS is conformally flat it seems likely the disturbance 
 would be on the light cone.  Then one might expect a null singularity 
 which converges to a point at the origin.  On the 
 other hand, the boundary field theory seems perfectly well behaved 
 and one would expect finite changes in the dilaton boundary value would lead 
 to finite energy states (e.g. scattered waves, black holes) in the 
 bulk.  We will see the second is true.  The fact that we 
 only get a finite reaction suggests this approach might lead to 
 a full quantum 
 description of matter dynamically collapsing to form a black 
 hole---a still unrealized goal.
 
 \indent It is also interesting to consider what happens 
 if one sends the gauge coupling in the boundary theory to zero.   While the 
 gauge theory becomes free the dilaton boundary value gets sent to 
 minus infinity and it seems very unlikely one would get a finite 
 response in the bulk.  Perhaps the simplest possibilty is a big crunch.  It would be very interesting to have a 
 full quantum gravity description of any cosmological singularity and to 
 be able to say something definitive about the possibility of a bounce 
 through a big crunch.
 
 \indent Analytic continuation between Euclidean and Lorentzian AdS 
 turns out to be significantly more subtle than is usually assumed 
 and in particular I will discuss the failure of a sensible analytic 
 continuation of the solution to Laplace's equation in either 
 direction.  I present a generic mode sum formalism for solving the 
 Lorentzian problem and then specialize to massless fields in AdS and 
 find several explicit solutions.  I then discuss how to estimate the 
 energy flowing through the boundary and in the bulk and why finite 
 changes in the dilaton boundary value result in finite energy states 
 in the bulk.  The disturbance created in the bulk by changing 
 boundary conditions is not confined to 
 the light cone and in the cases where it is large I estimate the size 
 of the black holes produced.  Finally, I mention an array of future 
 research possibilities.

 \setcounter{equation} {0}
 \section{The Euclidean solution and the failure of analytic continuation}

 \indent  Almost all of the work on string theory in AdS has been 
 done in the Euclidean signature with the assertion made that the (or 
 at least a) Lorentzian solution is given by an analytic continuation.   
 In this section I will show that the analytic continuation of the 
 Euclidean solution to Laplace's equation in AdS with time dependent 
 boundary conditions is not sensible.  In particular, one always 
 finds either branch cuts prevent a definition of the expressions in 
 question or complex results for given real boundary conditions.  The 
 solution for Euclidean $AdS_{d+1}$  for a
 minimally coupled massless scalar field in Poincar\'e 
 coordinates was given by Witten some time 
 ago~\cite{Wi}:
 \begin{equation}\label{WittenEuc}
 \Psi = c\int \mathrm{dx'^{d}} \underbrace{\frac{z^{d}}{(z^{2} + (x_{0} -
 x'_{0})^{2} +
 \sum_{i = 1}^{d-1} (x_{i} - x'_{i})^{2})^{d}}}_{G(\vec{x},\vec{x}',z)} 
 \, \cdot \,
 \phi (\vec{x}')
 \end{equation}
 where c is a constant, $\vec{x} = (x_{0},x_{1} \ldots, x_{d-1})$, $\phi =
 \Psi(z = 0)$ (i.e. the boundary value), $\nabla^{2}G(\vec{x},\vec{x}',z) =
 0$ for $z \ne 0$, and
 $G(\vec{x},\vec{x}',z) \rightarrow \delta(\vec{x} - \vec{x}')$ as $z
 \rightarrow 0$.  Recall the metric in Poincar\'e coordinates with Euclidean 
 signature is given by:
 \begin{equation}\label{AdSPoincaremetric}
\mathrm{ds}^{2} = \frac{\mathrm{d}z^{2} + \mathrm{d}x_{0}^{2} + 
\mathrm{d}x_{1}^{2} + \ldots{} +  \mathrm{d}x_{\mathrm{d} - 
1}^{2}}{z^{2}}   
\end{equation}

\begin{figure}
\begin{center}

\includegraphics[scale = .8]{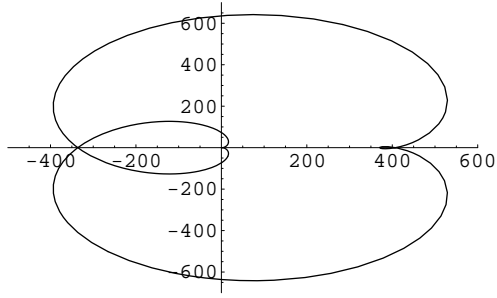}
    \caption{ $\epsilon_{1} = 1/10 +i/100$, $\, \, \epsilon_{2} = -1/10 +i/200$}
    \label{BrCut1}
\end{center} 
\end{figure}

\begin{figure}
\begin{center}
	\includegraphics[scale=.8]{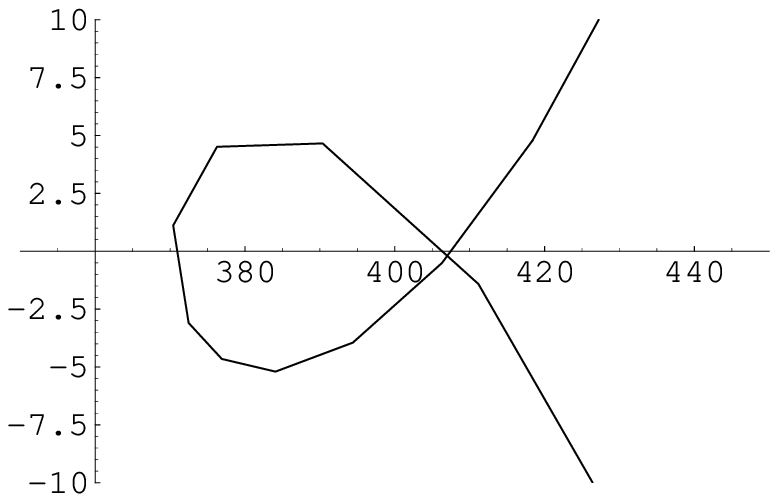}
	\caption{$\epsilon_{1} = 1/10 +i/100$, $\, \, \epsilon_{2} = -1/10 
	+i/200$ (detail)}
    \label{BrCut2}

    \end{center}
\end{figure}

\begin{figure}
\begin{center}
    \includegraphics[scale = .8]{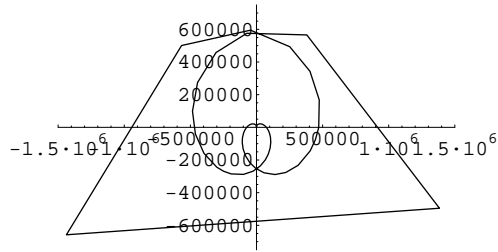}
    \caption{$\epsilon_{1} = 1/10 +i/5000$, $\, \, \epsilon_{2} = -1/1000 
	+i/10000$}
    \label{BrCut3}
\end{center}
\end{figure}

 \indent Before turning to the subtleties of continuing (\ref{WittenEuc}) let me first though remind 
 the reader of a well known subtlety in the transition from Euclidean 
 to Lorentzian space.  Namely, in Lorentzian space the bulk solution is not uniquely specified by
 Dirichlet boundary conditions;  one can always add normalizable
 solutions which vanish at the boundary.  As is very well known, the Laplace equation in Euclidean space with regular boundary
 conditions has a unique solution.  The assumption has been made in the 
 literature that to solve for scalar fields in 
 Lorentzian space one
 just adds the desired normalizable solution (determined by boundary
 conditions on a spacelike slice through AdS) to the analytically
 continued Euclidean solution~\cite{holoprobes} :
 \begin{equation}\label{EucContinued}
 \Psi_{A} = \Psi_{normalizable} + c'\int \mathrm{dx'^{d}}
 \frac{z^{d}}{(z^{2} - (x_{0} - x'_{0})^{2} +
 \sum_{i = 1}^{d-1} (x_{i} - x'_{i})^{2})^{d}} \, \cdot \,
 \phi (\vec{x}')
 \end{equation}
 Note there is now a pole in the denominator.  For the sake of simplicity
 I'll drop the normalizable
 piece and stick to spherical symmetry.  Moving the origin to ($x_{0}, x_{1}, 
 \ldots{}, x_{\mathrm{d} - 1}$) and introducing spherical
 coordinates $(\rho, \Omega)$ for the shifted $(x'_{1}, \ldots, x'_{d-1})$:
 \begin{equation}\label{SphericalAnalytic1}
 \Psi_{A} = c''\int_{-\infty}^{\infty} \mathrm{d\tau} \int_{0}^{\infty}
 \mathrm{d\rho} \, \frac{\rho^{d-2} z^{d}}{(z^{2} - (t - \tau)^{2} +
 \rho^{2})^{d}} \, \cdot \, \phi(\tau)
 \end{equation}
 Including a generic pole prescription,
 \begin{equation}\label{SphericalAnalytic2}
\Psi_{A} = c''\int_{-\infty}^{\infty}
 \mathrm{d\tilde{\tau}} \int_{0}^{\infty}
 \mathrm{d\tilde{\rho}} \, \frac{\tilde{\rho}^{d-2}}{(\tilde{\rho} -
 (\sqrt{\tilde{\tau}^{2} - 1} +i\epsilon_{1})^{d}(\tilde{\rho} -
 (-\sqrt{\tilde{\tau}^{2} - 1} +i\epsilon_{2})^{d}} \, \cdot \, \phi
 (z\tilde{\tau} + t)
 \end{equation}
 where I've defined rescaled variables $\tilde{\rho} = \frac{\rho}{z}$,
 $\tilde{\tau} = \frac{\tau - t}{z}$ and made a general distortion of the
 poles by \emph{complex} $\epsilon_{1}$, $\epsilon_{2}$ for doing the $\tilde{\rho}$
 integral first.  The problem comes from the
 $\sqrt{\tilde{\tau}^{2} - 1}$; this function has two branch points at
 $\tilde{\tau} = \pm 1$ and a branch cut between them.   The image of
 this branch cut in the $\tilde{\tau}$ integral may very well cross
 the real axis and hence make the $\tilde{\tau}$ integral undefined;
 the branch cut is moved off the real axis by the epsilon prescription
 and then wrapped around the origin by $\frac {1}{z^{n}}$.  Specifically
 for $AdS_{5}$, ($d = 4$), defining $v = \sqrt{\tilde{\tau}^{2} - 1}
 +i(\frac{\epsilon_{1} - \epsilon_{2}}{2})$ the integral
 over $\tilde{\rho}$ gives:
 \begin{equation}\label{exactresidue}
 -\frac{i(\epsilon_{1} + \epsilon_{2}) (52v^{2} + 15(\epsilon_{1} +
 \epsilon_{2})^{2})}{384v^{6}(v^{2} + \frac{(\epsilon_{1} +
 \epsilon_{2})^{2}}{4})} \, \, - \frac{1}{128}\frac{(v^{4} +
 \frac{3(\epsilon_{1} + \epsilon_{2})^{2}v^{2}}{2} + \frac{5(\epsilon_{1} +
 \epsilon_{2})^{4}}{16})\log{\frac{v + \frac{i(\epsilon_{1} +
 \epsilon_{2})}{2}}{-v + \frac{i(\epsilon_{1} +
 \epsilon_{2})}{2}}}}{v^{7}(v^{2} + \frac{(\epsilon_{1} +
 \epsilon_{2})^{2}}{4})}
 \end{equation}
 Along the branch cut $v = x + i(\frac{Real[\epsilon_{1} -
\epsilon_{2}]}{2})$
 where x $\in [ -1 + \frac{lm[\epsilon_{2} -
\epsilon_{1}]}{2}, 1 + \frac{lm[\epsilon_{2} -
\epsilon_{1}]}{2} ]$.  Expanding around the
 endpoints---$\vert x \vert \approx 1$---one finds
 (\ref{exactresidue}) $\approx \frac {-i\pi}{128x^{5}}$ and hence one
 endpoint is in the upper half plane and the other in the lower.  Then
 the branch cut crosses the real axis for any pole prescription and
 hence the $\tilde{\tau}$ integral isn't well defined.  
 Figures~\ref{BrCut1}~-~\ref{BrCut3} show typical branch cuts.

 \indent Now it is possible one might assert the above is wrong in the
 sense one should make an independent pole prescription  for the
$\tilde{\tau}$
 integral, despite the fact (\ref{exactresidue}) is well defined
 provided $Im(\epsilon_{1}) > 0, Im(\epsilon_{2}) > 0$ and $Im(\epsilon_{1})
 \neq Im(\epsilon_{2})$.  I will denote the new  distortions by complex $\gamma_{i}$.  Let us first examine the possibility that
 one first sends $\epsilon_{1}, \epsilon_{2} \rightarrow 0$:
 \begin{equation}\label{altresidue1}
 \int_{0}^{\infty} \mathrm{d\rho} (\qquad) = \frac
 {-i\pi}{128(\tilde{\tau}^{2} - 1)^{\frac{5}{2}}}
 \end{equation}
 and then makes a generic pole prescription
 \begin{eqnarray} \label{altresidue2}
 \int_{0}^{\infty} \mathrm{d\rho} (\qquad) = \frac {-i\pi}{128\Big( (\tilde{\tau} -
 (1+i\gamma_{1}))(\tilde{\tau} - 
 (-1+i\gamma_{2}))\Big)^{\frac{5}{2}}} \nonumber\\
 = \frac {-i\pi}{128\Big((\tilde{\tau} - i\frac{\gamma_{1} +
 \gamma_{2}}{2})^{2} -(1+i\frac{\gamma_{1} -
 \gamma_{2}}{2})^{2}\Big)^{\frac{5}{2}}}
 \end{eqnarray}
 The $\frac{5}{2}$ exponent means we have a branch cut between the points 
 $1+i\gamma_{1}$ and $-1+i\gamma_{2}$.  If these points are on the 
 same side of the real axis the expression is identically zero while if 
 they are on opposite sides the branch cut crosses the real axis. Again 
 we can't define the expression properly.  

 \indent On the other hand one might keep $\epsilon_{1}$ and
 $\epsilon_{2}$ finite and then make an independent pole prescription for
 the $\tilde{\tau}$ integral.  As it turns out, this does not help.  Defining $\mu = \frac{\tilde{\tau} -
 i\frac{\gamma_{1} + \gamma_{2}}{2}}{1+i\frac{\gamma_{1} - \gamma_{2}}{2}}$:
 \begin{equation}\label{altresidue4}
 \sqrt{\tilde{\tau}^{2} - 1} \rightarrow \pm \frac{1}{1+i\frac{\gamma_{1} -
 \gamma_{2}}{2}}\sqrt{\mu^{2} - 1}
 \end{equation}
 where the sign depends on which branch cut we take.  The differences
 from (\ref{exactresidue}) don't change the leading behavior and we
 again get a branch cut which crosses the real axis.  An additional
 independent pole prescription doesn't make the integral well defined.

\begin{figure}
    \begin{minipage} [b] {.5\linewidth}
\begin{picture} (0,0)
    	\put(10,-1){$\frac{\mathrm{Im}(\Psi)}{c''}$}
    \end{picture}

	\includegraphics[scale = .8]{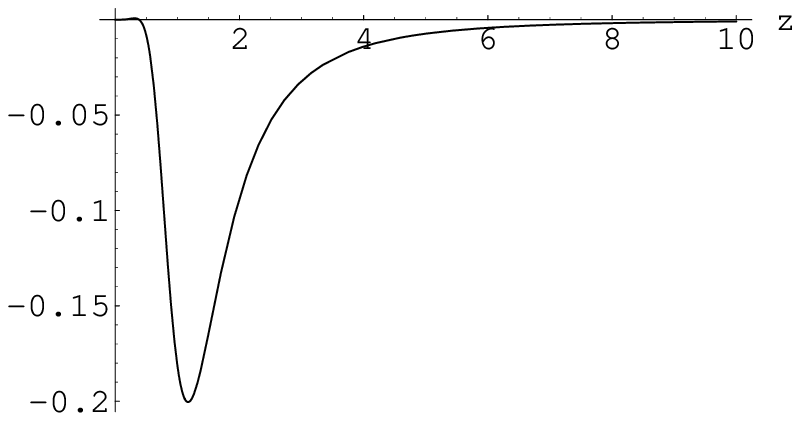}

\caption{}
\label{ImA}
\end{minipage} %
    \begin{minipage} [b] {.5\linewidth}
\begin{picture} (0,0)
    	\put(20,-2){$\frac{\mathrm{Im}(\Psi)}{c''}$}
    \end{picture}

	\includegraphics[scale=.8]{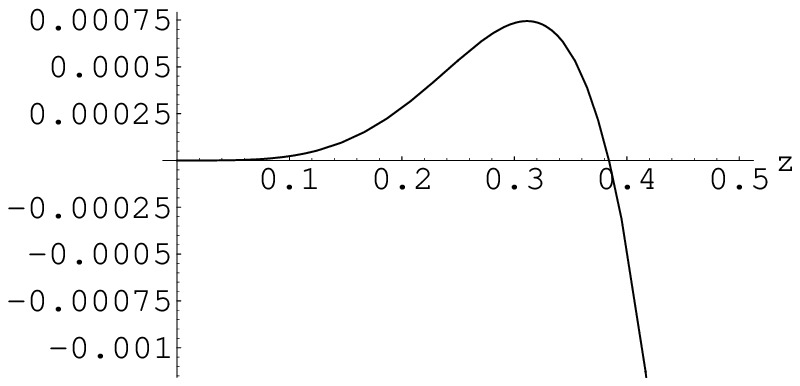}
	\caption{(detail of Fig. 4)}
    \label{ImB}

    \end{minipage}
\end{figure}

\begin{figure}
\vspace{.3in}
    \begin{center}
\begin{picture} (0,0)
    	\put(-85,0){$\frac{\mathrm{Im}(\Psi)}{c''}$}
    \end{picture}

	\includegraphics[scale = .8]{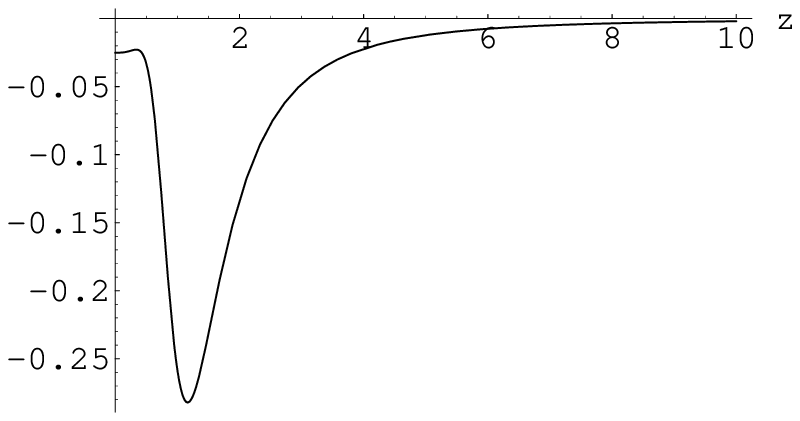}
\caption{}
	\label{ImC}
\end{center}
\end{figure}

\indent Hopefully by this point the reader is convinced that doing the
$\tilde{\rho}$ integral first is untenable.  Now let us consider doing the
$\tilde{\tau}$ integral first.  In particular, let us examine the ``solution'' for the
boundary function
\begin{equation}\label{continuedtime1st-1}
\phi = \frac{1}{(\tilde{\tau} - \mathrm{r}e^{i\theta})(\tilde{\tau} -
\mathrm{r}e^{-i\theta})} = \frac{1}{\tilde{\tau}^{2} + \mathrm{r}^{2} -
2\mathrm{r}\tau\cos{\theta}}
\end{equation}
where we take $0 < \theta < \frac{\pi}{2}$ or $\frac{\pi}{2} < \theta <
\pi$.  Neither this function nor its analytic continuation has a pole on
the real axis and both $\rightarrow 0$ as $\vert\tau\vert \rightarrow
\infty$.  We must make a pole prescription for the $\tau$ integral but 
after we do that integral we may safely take the $\rho$ integral to 
be real.  In all cases we get a complex solution for these real 
boundary conditions.  The explicit forms of resulting solutions are not terribly 
illuminating so 
I'll just plot the imaginary part of results for the case r~$ \, =1$ and $\theta = 
\frac{\pi}{4}$ at $t = 0.1$.  
Figures ~\ref{ImA}-~\ref{ImB} show the imaginary 
part of the result for both poles moved below the real axis and 
Figure~\ref{ImC} shows the result if there are poles 
moved into the second and fourth quadrants.  The imaginary parts of the 
solution for the remaining pole prescriptions, both poles up and 
poles in the first and third quadrants, are, respectively, the negative of the 
plotted results. 

\indent In all cases the analytic continuation does not 
produce a sensible result.  We actually never had a reason to expect that 
it necessarily would.  Algebraic
expressions generally continue without trouble (other than producing
divergences where there were none before, e.g. oscillatory expressions
(including mode solutions) to exponentials and vice versa).  If one has
integrands in which continuation produces a pole the resulting integral
has to be understood in the complex sense and that is considerably more
delicate than the corresponding real sense.  This is not, of course, to
say the continuation of all integrals and non-elementary functions is 
ill defined, but one should not blindly continue integral expressions or formalism.

\setcounter{equation} {0}
\section{Solving $\nabla^{2}\Psi = 0$ in Lorentzian $AdS_{d+1}$ with time
dependent boundary conditions}

\indent I now turn to solving Laplace's equation for massless scalar
fields in global coordinates.  The metric in these coordinates is given
by:
\begin{equation}\label{AdSmetric}
\mathrm{ds}^{2} = -\sec^{2}(\rho)\mathrm{d}\tau^{2} +
\sec^{2}(\rho)\mathrm{d}\rho^{2}
+\tan^{\mathrm{d-1}}(\rho)\mathrm{d}\Omega_{d-1}^{2}
\end{equation}
where $0\leq \rho \leq \frac {\pi}{2}$, $-\infty \leq \tau \leq \infty$,
and $\mathrm{d}\Omega_{d-1}^{2}$ is the metric for the unit 
(d-1)-sphere.  Note I've set the AdS radius to 1.  
It will be convenient to define $y = \sin^{2}{\rho}$.  In terms of 
these coordinates, the usual radial coordinate r $= \tan{\rho} = (\frac{1}{y} - 1)^{-\frac{1}{2}}$.  For simplicity I'll restrict my attention to spherical symmetry, but it is 
straightforward to add spherical harmonics to remove this restriction.  The mode solution at
frequency $\omega$ which is regular at the origin is given in terms of a
hypergeometric function F:
\begin{equation}\label{modesolution}
\chi(\mathrm{y}, \mathrm{t}, \omega) =
\mathrm{F}(-\frac{\omega}{2},\frac{\omega}{2}, \frac{\mathrm{d}}{2},
y) \, e^{-i\omega\mathrm{t}}
\end{equation}
This result is somewhat more transparent than previous results in the
literature~\cite{BulkBdy} as it has easily found limits:
\begin{equation}\label{hyperlimits}
\mathrm{F}(-\frac{\omega}{2},\frac{\omega}{2}, \frac{\mathrm{d}}{2}, 0) =
1 \qquad \mathrm{F}(-\frac{\omega}{2},\frac{\omega}{2},
\frac{\mathrm{d}}{2}, 1) =
\frac{\Gamma(\frac{\mathrm{d}}{2})^{2}}{\Gamma(\frac{\mathrm{d}+
\omega}{2})\Gamma(\frac{\mathrm{d}-\omega}{2})}
\end{equation}

\indent The fact that the parameters of a hypergeometric function 
involve the frequency $\omega$ of the mode makes extracting usable
expressions from a mode sum formalism technically a bit difficult.  Using
Poincar\'e coordinates is potentially quite problematic because of the 
coordinate horizon; it's not entirely clear how to define the analog 
of a regular solution since the usual 
Bessel function solutions (including the part usually identified as 
the normalizable mode) oscillate with unbounded magnitude near the 
horizon.  I'll stick to global coordinates.  The
position space solution, if it could be found, might be simpler although
in practice even the Witten Euclidean position space solution tends to be
technically hard to work with analytically.  One can show with a bit of uninspiring algebra
that the solution in the Lorentzian case that one might expect---some function with delta
function support on the light cone--actually is not a solution except in
$AdS_2$.  I will be content here with a mode sum
solution.

\indent First I will describe some general formalism for any 
spherically symmetric solution (not just massless fields in AdS) if one has a complete set of mode solutions $\chi(\omega,
r) \, e^{-i\omega\mathrm{t}}$ and seeks a solution subject to boundary conditions on a surface at $\Psi(r =
\mathrm{r}_{\mathrm{b}},\mathrm{t}) = \phi(\mathrm{t})$.  For massless 
fields one may safely impose boundary conditions at infinity, although 
for the massive case one needs (apparently) to impose a cutoff and 
put boundary conditions at some finite radius.  It is again 
straightforward to extend the formalism to the non-spherically 
symmetric case.  The advertised solution is
\begin{equation}\label{genericsolution1}
\Psi(\mathrm{r,t}) = \frac{1}{2\pi}\int_{-\infty}^{\infty}
 \mathrm{d\omega} \int_{-\infty}^{\infty}
 \mathrm{d\tau} \, \, e^{i\omega(\tau - \mathrm{t})}
\, \frac{\chi(\omega,\mathrm{r})}{\chi(\omega,\mathrm{r}_{\mathrm{b}})} 
\, \phi(\tau)
\end{equation}
One can tell this is right as follows:
\begin{equation}\label{genericsolution2}
\Psi(\mathrm{r}_{\mathrm{b}},\mathrm{t}) = \int_{-\infty}^{\infty}
 \mathrm{d\tau} \underbrace{\frac{1}{2\pi} \int_{-\infty}^{\infty}
 \mathrm{d\omega} \, \, e^{i\omega(\tau - \mathrm{t})}}_{\delta(\tau -
\mathrm{t})} \, \phi(\tau) \quad = \, \, \phi(\mathrm{t})
\end{equation}
as required and
\begin{equation}\label{genericsolution3}
\Psi(\mathrm{r}, \mathrm{t}) = \int_{-\infty}^{\infty}
 \mathrm{d\omega} \,
\underbrace{\frac{1}{2\pi\chi(\omega,\mathrm{r}_{\mathrm{b}})}
\int_{-\infty}^{\infty}
 \mathrm{d\tau} \, \,
\phi(\tau)e^{i\omega\tau}}_{\mathrm{f(\omega)}}\chi(\omega,\mathrm{r}) 
\, e^{-i\omega\mathrm{t}}
\end{equation}
i.e. $\Psi$ is just a superposition of mode solutions and hence a
solution.

\indent Generically there are poles at frequencies where
$\chi(\omega,\mathrm{r}_{\mathrm{b}}) = 0$, i.e. the frequencies of the
normalizable
modes.  I will only be concerned here with massless fields, 
although I believe this is a sensible definition of normalizable 
modes for theories with a cutoff.  Then I distort $\omega \rightarrow \omega + i\epsilon(\omega)$ for
$\epsilon(\omega)$ a smooth function which
goes to a nonzero constant in the neighborhood of each pole and is
small compared to any relevant scale.  Then,
\begin{equation}\label{distortedformula1}
\Psi(\mathrm{r,t}) = \frac{1}{2\pi}\int_{-\infty}^{\infty}
 \mathrm{d\tau}\int \mathrm{d\omega} \, \, e^{i\omega(\tau - 
 \mathrm{t})} \, 
\frac{\chi(\omega + i\epsilon(\omega), \mathrm{r})}{\chi(\omega +
 i\epsilon(\omega), \mathrm{r}_{\mathrm{b}})}\, \phi(\tau)
 \end{equation}
 Note I've been careful to preserve the pole structure of the
 integrand.  Assuming $\chi(\mathrm{r}_{\mathrm{b}}, \omega)$ has at
 most a countable set of zeroes $\{\omega_{\mathrm{n}}\}$ (this is
 certainly true for AdS) and defining
 $\epsilon(\omega_{\mathrm{n}}) = \epsilon_{\mathrm{n}}:$
 \begin{eqnarray}\label{distortedformula2}
 \Psi(\mathrm{r,t}) = -i\int_{-\infty}^{\mathrm{t}}
 \mathrm{d\tau} \, \, \phi(\tau) \sum_{\mathrm{n} \in \mathbf{Z}, \;
\epsilon_{\mathrm{n}} > 0 } \, e^{i\omega(\tau - \mathrm{t})}
 \, \mathrm{Res}(\frac{\chi(\omega + i\epsilon(\omega),
 \mathrm{r})}{\chi(\omega + i\epsilon(\omega),
 \mathrm{r}_{\mathrm{b}})}) \Bigg\arrowvert_{\omega = \omega_{\mathrm{n}} -
 i\epsilon_{\mathrm{n}}} \nonumber\\
+ i\int_{\mathrm{t}}^{\infty}
 \mathrm{d\tau} \phi(\tau) \sum_{\mathrm{n} \in \mathbf{Z}, \;
\epsilon_{\mathrm{n}} < 0 } e^{i\omega(\tau - \mathrm{t})}
 \, \mathrm{Res}(\frac{\chi(\omega + i\epsilon(\omega),
 \mathrm{r})}{\chi(\omega + i\epsilon(\omega),
 \mathrm{r}_{\mathrm{b}})}) \Bigg\arrowvert_{\omega = \omega_{\mathrm{n}} -
i\epsilon_{\mathrm{n}}}
\end{eqnarray}
where the poles are now at $\omega = \omega_{\mathrm{n}} -
i\epsilon_{\mathrm{n}}$.   Note then $\Psi$ breaks up into an
advanced and retarded piece and we have a very natural way to select
the desired propagator from among this countably infinite set---namely
only the propagator with each pole moved down is casual.  This is, of
course, the retarded propagator.

\indent Returning to the case of massless fields in AdS, for 
$AdS_{5}$~$ \, \omega_{\mathrm{n}} = 2\mathrm{n}$ and
\begin{equation}\label{AdS5retarded}
\Psi(\mathrm{y, t})_{\mathrm{Retarded}} = 2i\sum_{\mathrm{n} \in
\mathbf{Z}} (-1)^{\mathrm{n}} \mathrm{n} (\mathrm{n}^{2} - 1) \,
\mathrm{F(-n,n,2,y)} \int_{-\infty}^{\mathrm{t}}
 \mathrm{d\tau} \, \, \phi(\tau) \, e^{(2i\mathrm{n} + \epsilon)(\tau - \mathrm{t})}
\end{equation}
where $\epsilon_{\mathrm{n}} = \epsilon > 0 \quad \forall \ \mathrm{n}$.

\indent Now that we have a Lorentzian solution the reader might
wonder, despite the previous section, whether there is a sensible analtic
continuation for any pole prescription.  The short answer is apparently
no.  Consider rotating the contour by angle $\theta > 0$ in either the 
clockwise or counterclockwise direction.  Then define the 
Euclidean time $t_{E} = \pm e^{\mp i \theta} t$ where the upper and 
lower signs refer to the counterclockwise and clockwise directions 
respectively.  We immediately get a divergence unless we take:
\begin{displaymath}
\epsilon_{\mathrm{n}} = \Bigg\{ \begin{array}{ll}
\mp \epsilon \quad \quad $ if n $>$ 0 $ \\
\pm \epsilon \quad \quad $ if n $<$ 0 $
\end{array}
    \end{displaymath}

where $\epsilon > 0$---i.e. a Feynman
prescription.  Once we rotate the contour by angle $\theta \neq 0$ can use the hypergeometric function generating 
function(\cite{Bateman}) to do the
infinite sum with the result:
\begin{eqnarray}\label{Feynmanresult}
\Psi(\mathrm{y,t})_{\mathrm{F}} = -\int_{-\infty}^{0}
\mathrm{d}t_{E} \, \frac{(\mathrm{analytic \, \, function})}{(1 + 
2(1-2y)e^{\mp2ie^{\pm i \theta}t_{E}} + e^{\mp4ie^{\pm i 
\theta}t_{E}})^{\frac{5}{2}}} \nonumber\\
-\int_{0}^{\infty}
\mathrm{d}t_{E} \, \frac{(\mathrm{analytic \, \, function})}{(1 + 
2(1-2y)e^{\pm2ie^{\pm i \theta}t_{E}} + e^{\pm4ie^{\pm i 
\theta}t_{E}})^{\frac{5}{2}}}\end{eqnarray}
The fractional power means we again have a branch cut to worry about.  
Specifically as we change $t_{E}$ the contents of the parenthesis in 
the denominator in each of the above 
terms---a complex number even for real $t_{E}$---goes through this 
branch cut.  Working out the details one finds for each integral given any y within the 
range $\frac{1}{2} < y < 1$ and any point on the branch cut, there is a 
countably infinite set of rotation angles $\{ \theta_{\mathrm{n}} \}$ 
between 0 and $\frac{\pi}{2}$ and associated times  $\{ 
t_{{E}_{\mathrm{n}}} \}$ when the expression in the denominator moves through the branch cut 
at that point.  There is a similiar, albeit slightly more complicated, 
story for $0 < y < \frac{1}{2}$. The above is, however, sufficient to show 
the failure of the continuation.

\setcounter{equation} {0}
\section{Examples}

\indent Now let us examine some specific solutions.  If one takes $\phi =
\mathrm{A}e^{\mathrm{at}}$ (Real(a) $\geq$ $0$) it is possible to
explicitly do the sum and integral in (\ref{AdS5retarded}) with the
result
\begin{equation}\label{exponeqn1}
\Psi =
\mathrm{A} \, \frac{\mathrm{F}(\frac{i\mathrm{a}}{2},-\frac{i\mathrm{a}}{2},
2, \mathrm{y})}{\mathrm{F}(\frac{i\mathrm{a}}{2},-\frac{i\mathrm{a}}{2},
2, 1)} \, e^{\mathrm{at}} = \mathrm{A}
\, 
\frac{\frac{\mathrm{a}\pi}{2}(1+\frac{\mathrm{a}^2}{4})}{\sinh(\frac{\mathrm{a}\pi}{2})} \,
\mathrm{F}(\frac{i\mathrm{a}}{2},-\frac{i\mathrm{a}}{2}, 2,
\mathrm{y}) \, e^{\mathrm{at}}
\end{equation}
which also may be written with a $= i\omega$ as
\begin{equation}\label{exponeqn2}
\Psi = \mathrm{A} \,
\frac{\frac{\omega\pi}{2}(1+\frac{\omega^2}{4})}{\sin(\frac{\omega\pi}{2})}
 \, \mathrm{F}(\frac{\omega}{2},-\frac{\omega}{2}, 2,
\mathrm{y}) \, e^{i\omega\mathrm{t}}
\end{equation}
There are several points to note here.  For $a$ = 0 we find for $\phi$ = 
c, where c is some constant, $\Psi$ = c.  Of course, this had to be.  We also recover
the mode solutions for real $\omega$.  As $\omega \rightarrow 2\mathrm{n}$
for n $\in \mathbf{Z}^{\setminus \{0,\pm1\}}$ one is forcing a mode to
oscillate which approaches zero near the boundary and hence gets an
arbitrarily large response in the bulk.

\begin{figure}
    \begin{minipage} [b] {.5\linewidth}
\includegraphics[scale = .8]{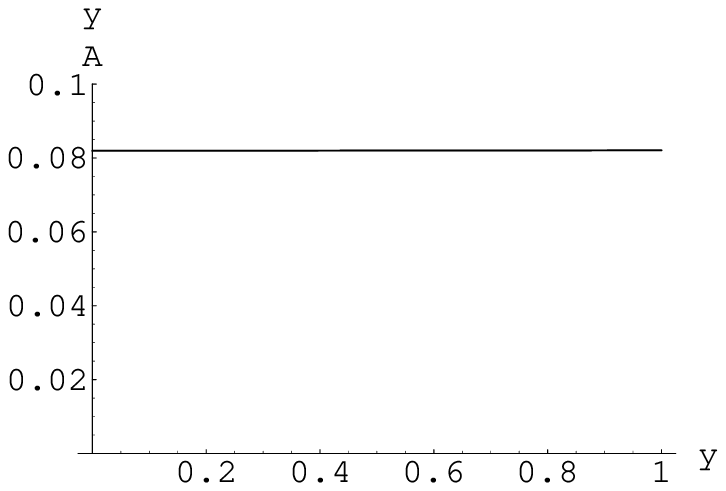}
    \caption{ a = 0.1, t = -25}
    \label{figexp1}
    \end{minipage} %
    \begin{minipage} [b] {.5\linewidth}
	\includegraphics[scale=.8]{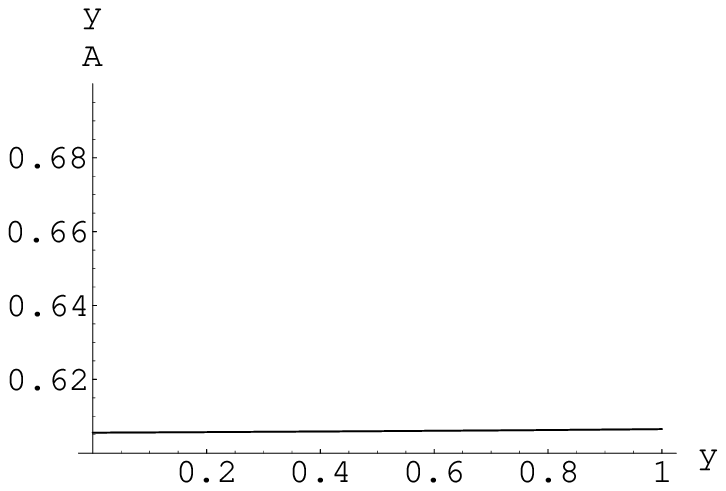}
	\caption{a = 0.1, t = -5}
    \label{figexp2}

    \end{minipage}
\end{figure}

\begin{figure}
\begin{center}
    \includegraphics[scale = .8]{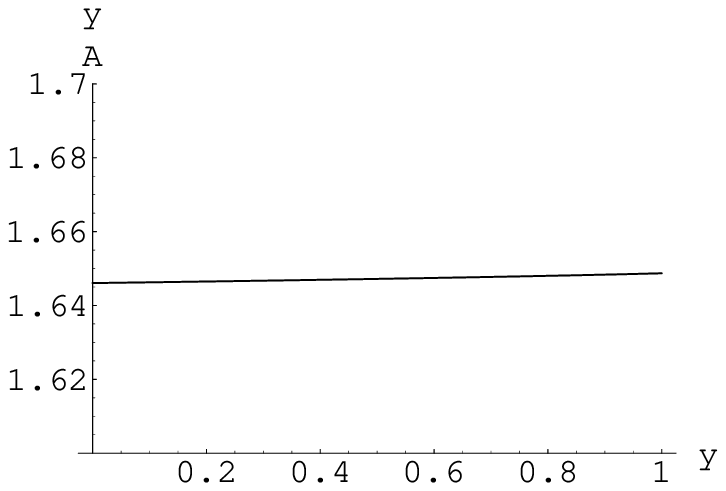}
    \caption{ a = 0.1, t = 5}
    \label{figexp3}
\end{center}

\end{figure}

\begin{figure}
    \begin{minipage} [b] {.5\linewidth}
\includegraphics[scale = .8]{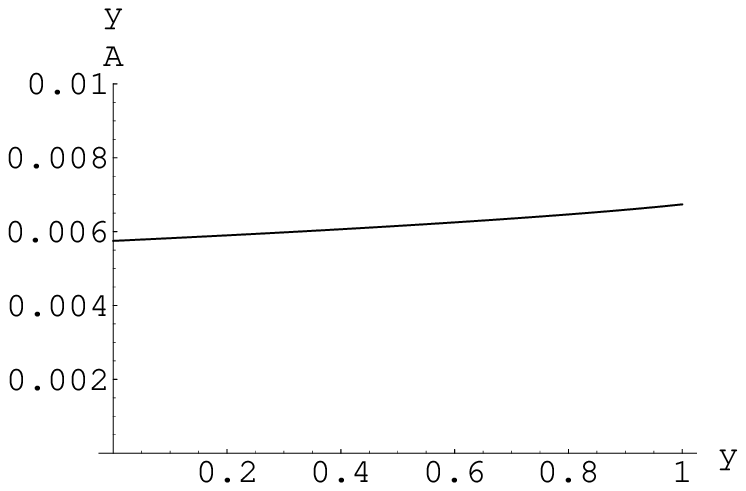}
    \caption{ a = 1, t = -5}
    \label{figexp4}

\end{minipage} %
    \begin{minipage} [b] {.5\linewidth}
	\includegraphics[scale=.8]{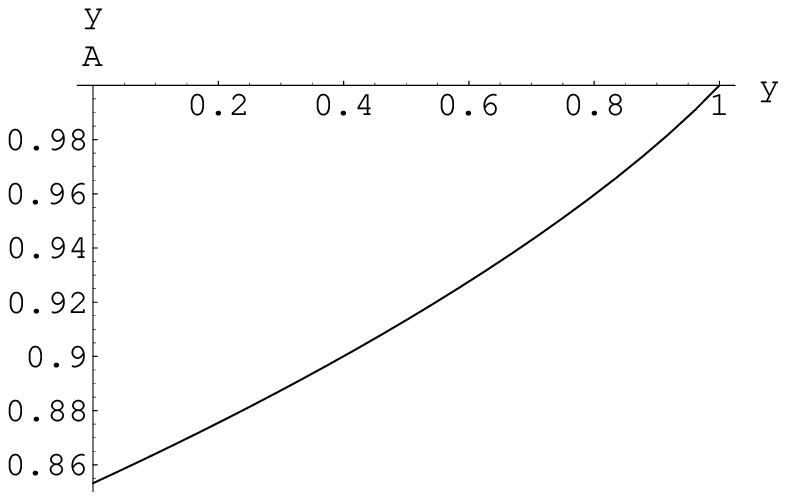}
	\caption{a = 1, t = 0}
    \label{figexp5}

    \end{minipage}
\end{figure}

\begin{figure}
    \begin{minipage} [b] {.5\linewidth}
\includegraphics[scale = .8]{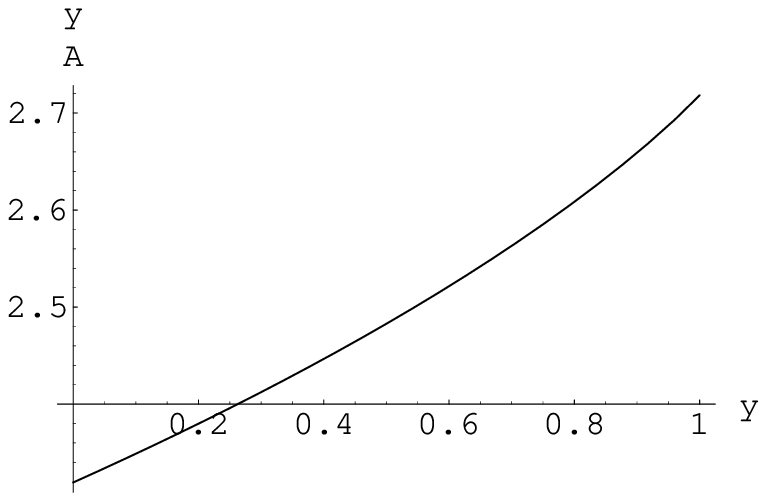}
    \caption{ a = 1, t = 1}
    \label{figexp6}

\end{minipage} %
    \begin{minipage} [b] {.5\linewidth}
	\includegraphics[scale=.8]{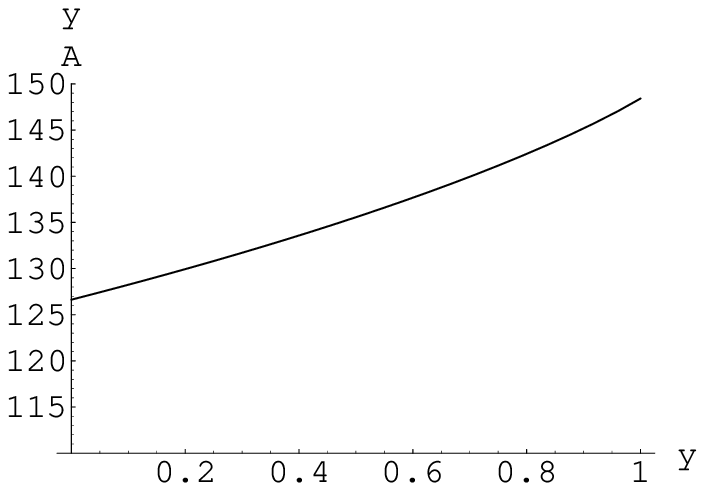}
	\caption{a = 1, t = 5}
    \label{figexp7}

    \end{minipage}
\end{figure}

\begin{figure}
    \begin{minipage} [b] {.5\linewidth}
\includegraphics[scale = .8]{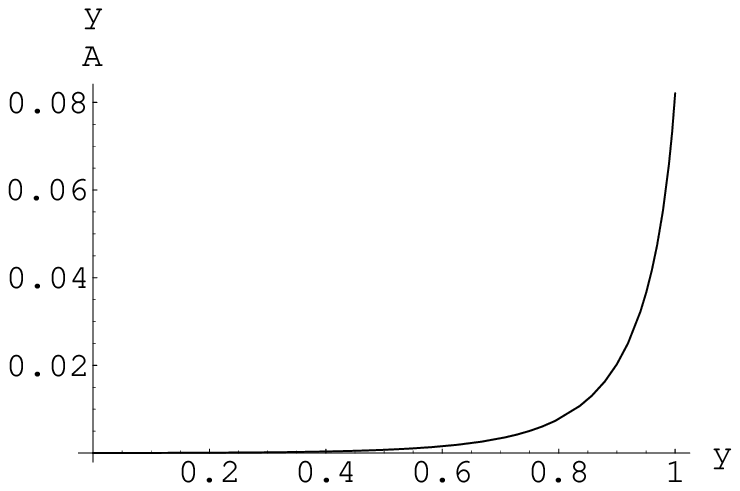}
    \caption{ a = 10, t = -0.25}
    \label{figexp8}

\end{minipage} %
    \begin{minipage} [b] {.5\linewidth}
	\includegraphics[scale=.8]{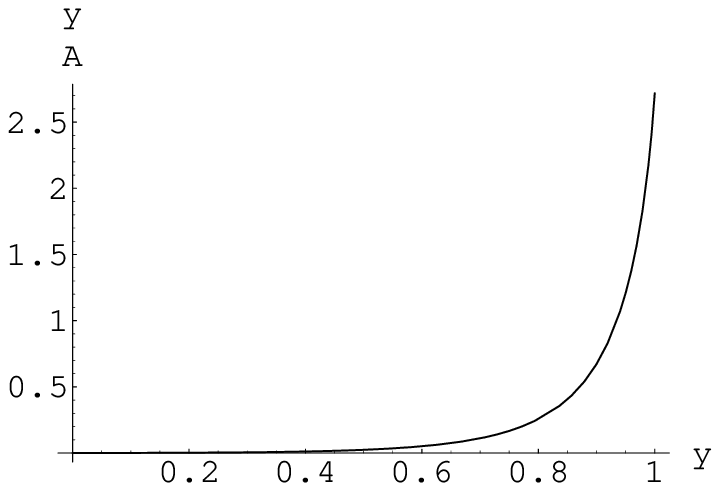}
	\caption{a = 10, t = 0.1}
    \label{figexp10}

    \end{minipage}
\end{figure}

\begin{figure}
    \begin{minipage} [b] {.5\linewidth}
\includegraphics[scale = .8]{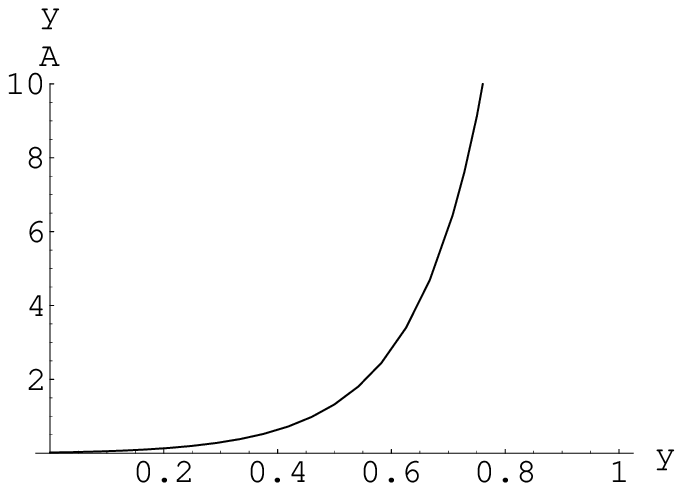}
    \caption{ a = 10, t = 0.5}
    \label{figexp11}

\end{minipage} %
    \begin{minipage} [b] {.5\linewidth}
	\includegraphics[scale=.8]{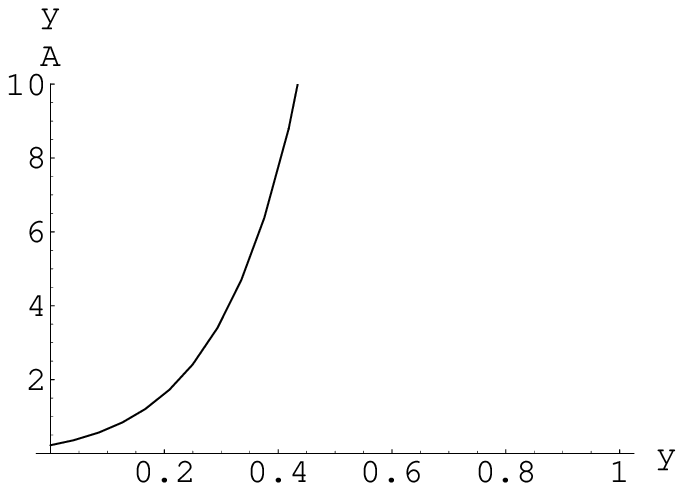}
	\caption{a = 10, t = 0.75}
    \label{figexp12}

    \end{minipage}
\end{figure}

\indent Upon first examination one notices (\ref{exponeqn2}) is factorized into
time dependent and spatial dependent pieces and hence might wonder whether
this solution is in fact retarded.  Of course one knows of a simple
example that is retarded and factorizes this way---$\Psi =
e^{i\mathrm{kx}}e^{-i\omega\mathrm{t}}$.  
We do in fact have retardation; figures~\ref{figexp1}~-~\ref{figexp12}~ plot solutions with $\phi = 
e^{\mathrm{at}} $ for real $a$ and $A = 1$.  If we change
the boundary conditions much more slowly than the AdS time scale---set to 1
here---the situation is quasi-static and the bulk solution is nearly
homogeneous.  If we change the boundary conditions much faster than 
the AdS time scale the field near the boundary changes faster than 
signals can travel to points near the origin and we get a very 
inhomogeneous solution.

\indent For most boundary conditions obtaining an exact and explicit solution is 
extremely difficult.  However, relatively recently an asymptotic series for 
the relevant hypergeometric function for $\vert \mathrm{n} \vert \gg 
1$ has been given in terms of Bessel functions~\cite{Jones}.  The Bessel functions can then be expanded in the usual 
asymptotic series provided one is content to confine one's attention 
to points a bit away from the origin (here I will get approximations 
reliable for y $\geq 0.01$).  Then provided the integral in (\ref{AdS5retarded}) 
can be done and gives a relatively simple result, we can get a reliable 
approximation for $\Psi$.  Note that, however, the sum does not 
generically converge quickly and hence it is important to include 
terms of arbitrarily large order.  In particular, one can get a good 
approximation by truncating the asymptotic 
expansion to some finite number of terms (I'll take three) and then doing 
the infinite sums.  

\begin{figure}
 \vspace{.5in}
    \begin{minipage} [b] {.5\linewidth}
\begin{picture} (0,0)
    	\put(20,120){$\frac{\Psi}{\mathrm{A}}$}
	\put(180,50){y}
    \end{picture}
	\includegraphics[scale = .8]{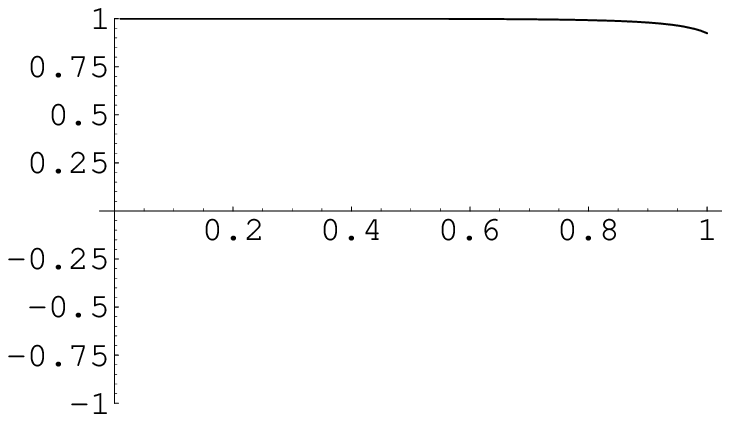}
    \caption{ a = 10, t = -0.25}
    \label{tanh5}

\end{minipage} %
    \begin{minipage} [b] {.5\linewidth}
\begin{picture} (0,0)
    	\put(10,10){$\frac{\Psi}{\mathrm{A}}$}
	\put(175,-10){y}
    \end{picture}

	\includegraphics[scale=.8]{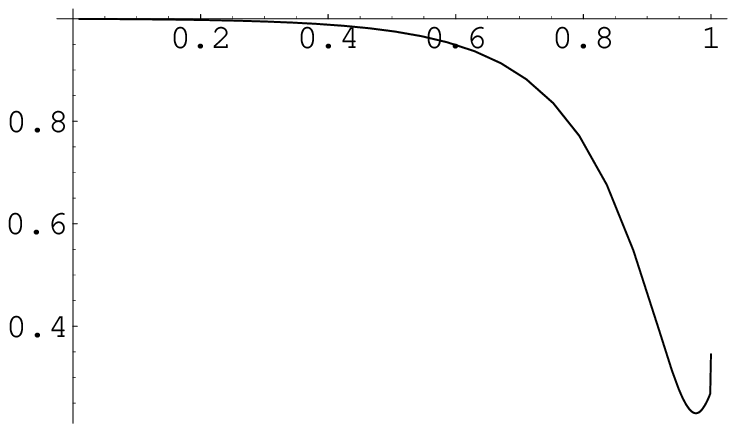}
	\caption{a = 10, t = 0.1}
    \label{tanh6}

    \end{minipage}
\end{figure}

\begin{figure}
 \vspace{.5in}
    \begin{minipage} [b] {.5\linewidth}
\begin{picture} (0,0)
    	\put(20,120){$\frac{\Psi}{\mathrm{A}}$}
	\put(182,20){y}
    \end{picture}
	\includegraphics[scale = .8]{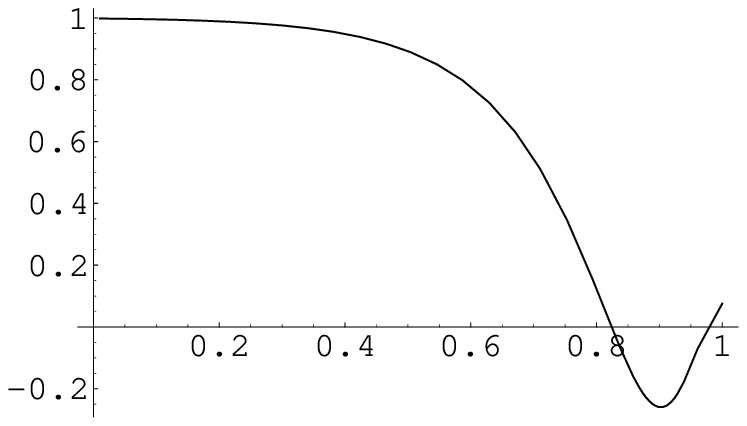}
    \caption{ a = 10, t = 0.25}
    \label{tanh7}

\end{minipage} %
    \begin{minipage} [b] {.5\linewidth}
\begin{picture} (0,0)
    	\put(15,10){$\frac{\Psi}{\mathrm{A}}$}
	\put(175,-35){y}
    \end{picture}

	\includegraphics[scale=.8]{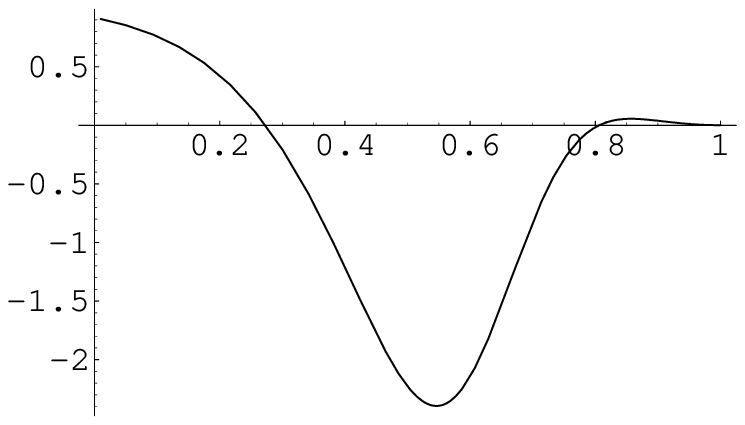}
	\caption{a = 10, t = 0.65}
    \label{tanh8}

    \end{minipage}
\end{figure}

\begin{figure}
 \vspace{.5in}
    \begin{minipage} [b] {.5\linewidth}
\begin{picture} (0,0)
    	\put(12,120){$\frac{\Psi}{\mathrm{A}}$}
	\put(178,85){y}
    \end{picture}
	\includegraphics[scale = .8]{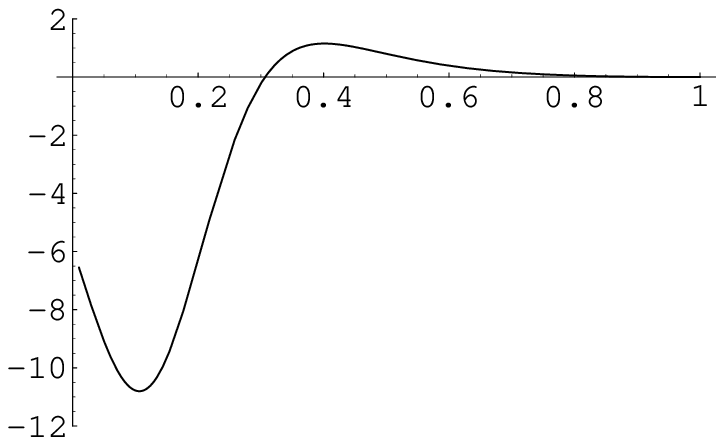}
    \caption{ a = 10, t = 1.1}
    \label{tanh10}

\end{minipage} %
    \begin{minipage} [b] {.5\linewidth}
\begin{picture} (0,0)
    	\put(10,10){$\frac{\Psi}{\mathrm{A}}$}
	\put(175,-20){y}
    \end{picture}

	\includegraphics[scale=.8]{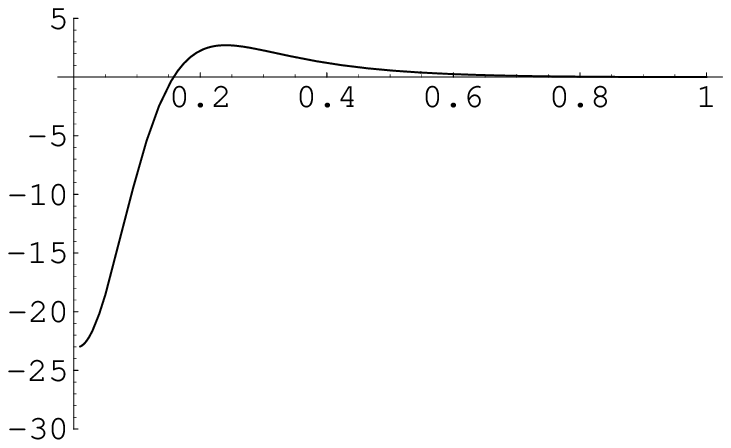}
	\caption{a = 10, t = 1.25}
    \label{tanh11}

    \end{minipage}
\end{figure}

\begin{figure}
 \vspace{.5in}
    \begin{minipage} [b] {.5\linewidth}
\begin{picture} (0,0)
    	\put(15,120){$\frac{\Psi}{\mathrm{A}}$}
	\put(182,60){y}
    \end{picture}
	\includegraphics[scale = .8]{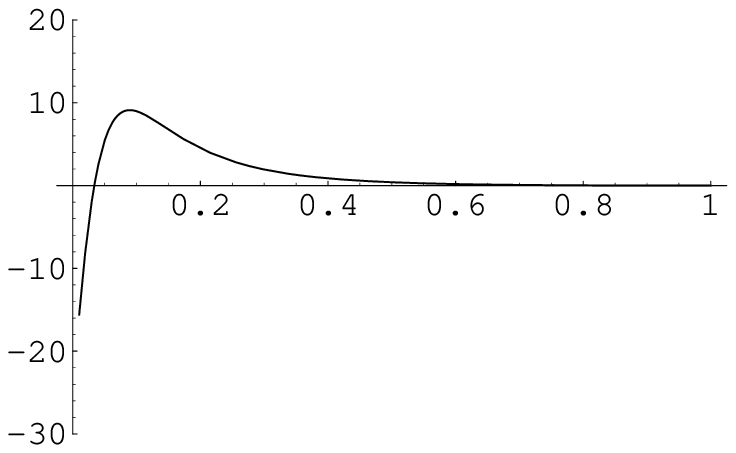}
    \caption{ a = 10, t = 1.4}
    \label{tanh12}

\end{minipage} %
    \begin{minipage} [b] {.5\linewidth}
\begin{picture} (0,0)
    	\put(6,10){$\frac{\Psi}{\mathrm{A}}$}
	\put(175,-92){y}
    \end{picture}

	\includegraphics[scale=.8]{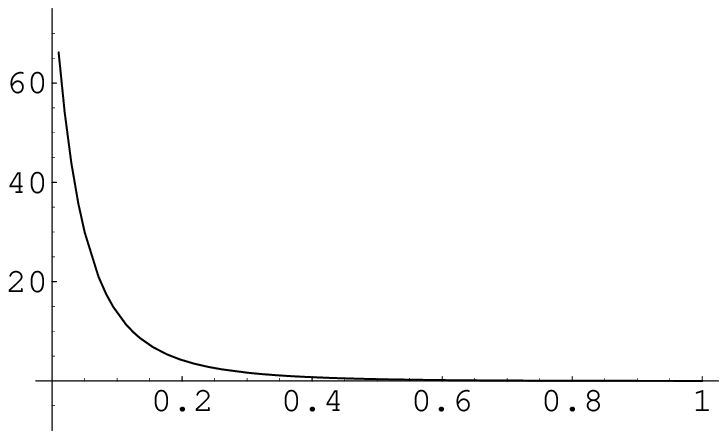}
	\caption{a = 10, t = 1.6}
    \label{tanh13}

    \end{minipage}
\end{figure}

\begin{figure}
 \vspace{.5in}
    \begin{minipage} [b] {.5\linewidth}
\begin{picture} (0,0)
    	\put(10,120){$\frac{\Psi}{\mathrm{A}}$}
	\put(175,65){t}
    \end{picture}
	\includegraphics[scale = .8]{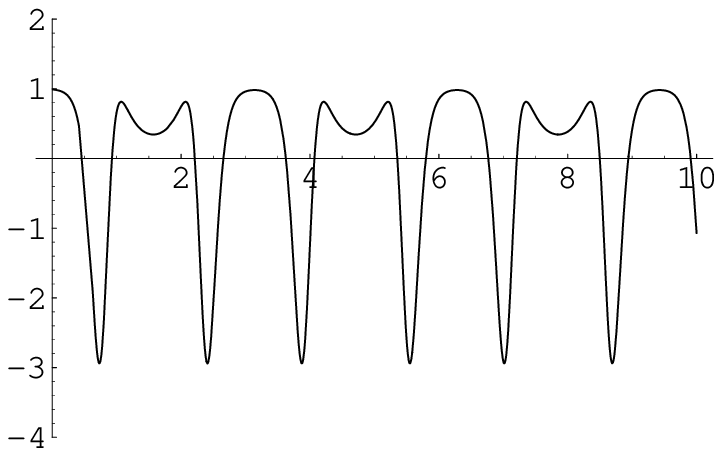}
    \caption{ a = 10, y = 0.5}
    \label{tanh16}

\end{minipage} %
    \begin{minipage} [b] {.5\linewidth}
\begin{picture} (0,0)
    	\put(10,10){$\frac{\Psi}{\mathrm{A}}$}
	\put(175,-65){t}
    \end{picture}

	\includegraphics[scale=.8]{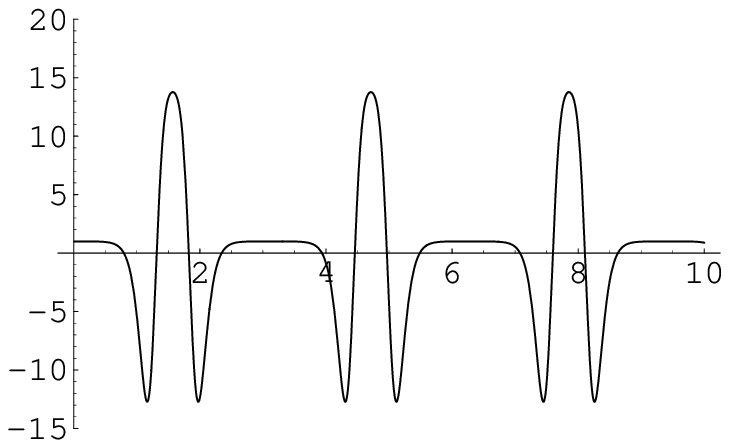}
	\caption{a = 10, y = 0.1}
    \label{tanh18}

    \end{minipage}
\end{figure}

\indent Consider the finite perturbation 
\begin{equation}
\phi = \frac{\mathrm{A}}{e^{\mathrm{at}} + \mathrm{c}} = 
\frac{\mathrm{A}}{\mathrm{c}} \Bigg( \frac{1}{1 + e^{\mathrm{a}(\mathrm{t} - 
\frac{\log{\mathrm{c}}}{\mathrm{a}})}} \Bigg) = 
\frac{\mathrm{A}}{2\mathrm{c}}\Bigg(1 - 
\tanh\Bigg(\frac{\mathrm{a}}{2}\Big(\mathrm{t} - 
\frac{\log{\mathrm{c}}}{\mathrm{a}}\Big)\Bigg)\Bigg)
\end{equation}
for c $> 0$, a $> 0$.  This is a smooth transition between 
$\frac{\mathrm{A}}{\mathrm{c}}$ and zero.  One finds for 
$e^{\mathrm{at}} < \mathrm{c}$
\begin{equation}\label{finiteeqn1}
\Psi = \frac{\mathrm{A}}{\mathrm{c}} + \frac{\mathrm{A} 
\pi}{\mathrm{c}}\sum_{\mathrm{k} = 1}^{\infty} \Bigg( 
-\frac{e^{\mathrm{at}}}{\mathrm{c}}\Bigg)^{\mathrm{k}} \, \, \frac{\frac{\mathrm{ak}}{2}(1+\frac{(\mathrm{ak})^2}{4})}{\sinh(\frac{\pi \mathrm{ak}}{2})}
\, \mathrm{F}(\frac{i\mathrm{ak}}{2},-\frac{i\mathrm{ak}}{2},2, \mathrm{y})
\end{equation}
and for $e^{\mathrm{at}} > \mathrm{c}$
\begin{eqnarray}\label{finiteeqn2}
\Psi = \frac{2\pi \mathrm{A}}{\mathrm{ac}} \sum_{\mathrm{n} \in 
\mathbf{Z}, \, \mathrm{n} \neq 0, \pm 1} 
\frac{(-1)^{\mathrm{n}} \, \mathrm{n}(\mathrm{n}^{2} - 
1) \, \mathrm{F}(-\mathrm{n}, \mathrm{n}, 2, 
\mathrm{y})}{\sinh(\frac{2\pi 
\mathrm{n}}{\mathrm{a}})} \, e^{-2i\mathrm{n}(\mathrm{t} - 
\frac{\log{\mathrm{c}}}{\mathrm{a}})} \nonumber\\
-\frac{\mathrm{A} \pi}{\mathrm{c}}\sum_{\mathrm{k} = 1}^{\infty} \Bigg( 
-\frac{\mathrm{c}}{e^{\mathrm{at}}}\Bigg)^{\mathrm{k}}\, \, \frac{\frac{\mathrm{ak}}{2}(1+\frac{(\mathrm{ak})^2}{4})}{\sinh(\frac{\pi \mathrm{ak}}{2})}
\, \mathrm{F}(\frac{i\mathrm{ak}}{2},-\frac{i\mathrm{ak}}{2},\mathrm{y})
\end{eqnarray}
Expression (\ref{finiteeqn1}) and the second piece of (\ref{finiteeqn2}) 
smoothly match onto the boundary value.  The first term in 
(\ref{finiteeqn2}) is a sum of undamped normalizable modes.  It is exponentially supressed 
if $a$ is small but if $a$ is large modes with frequencies up to 
$ \approx $ $a$ make a significant contribution.  This is exactly 
what one expects physically; if we change the boundary conditions 
slowly we have a quasi-static, nearly homogeneous solution, but if we 
change them quickly by the time disturbances could propagate across 
AdS the boundary conditions become nearly static and hence nearly 
reflective and at late times we end up with waves which bounce back 
and forth indefinitely.  Figures~\ref{tanh5}~-~\ref{tanh18} 
are plots with $a = 10$ using the approximations described above and are 
expected to 
be accurate up to $0.1$ percent.   
As one increases $a$, many more normalizable modes make significant 
contributions and the resulting undamped oscillations are greater.  In 
particular, these oscillations are smaller by nearly an order of 
magnitude if we take $a = 5$ instead of $a = 10$.

\begin{figure}
 \vspace{.5in}
    \begin{minipage} [b] {.5\linewidth}
\begin{picture} (0,0)
    	\put(21,105){$\frac{\Psi}{\mathrm{A}}$}
	\put(165,7){y}
    \end{picture}
	\includegraphics[scale = .8]{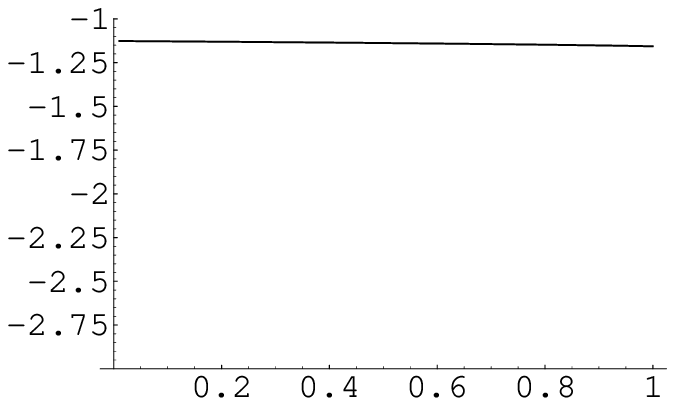}
    \caption{ a = 1, t = -2}
    \label{bigcrnch1}

\end{minipage} %
    \begin{minipage} [b] {.5\linewidth}
\begin{picture} (0,0)
    	\put(18,6){$\frac{\Psi}{\mathrm{A}}$}
	\put(165,-93){y}
    \end{picture}

	\includegraphics[scale=.8]{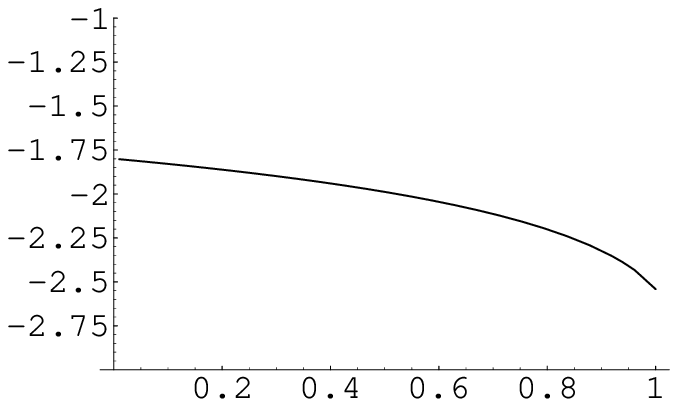}
	\caption{a = 1, t = -0.5}
    \label{bigcrnch2}

    \end{minipage}
\end{figure}

\begin{figure}
 \vspace{.5in}
    \begin{minipage} [b] {.5\linewidth}
\begin{picture} (0,0)
    	\put(20,100){$\frac{\Psi}{\mathrm{A}}$}
	\put(170,20){y}
    \end{picture}
	\includegraphics[scale = .8]{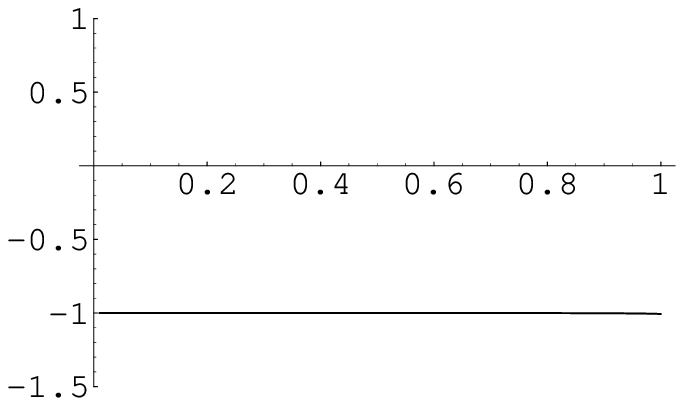}
    \caption{ a = 10, t = -0.5}
    \label{bigcrnch3}

\end{minipage} %
    \begin{minipage} [b] {.5\linewidth}
\begin{picture} (0,0)
    	\put(15,10){$\frac{\Psi}{\mathrm{A}}$}
	\put(175,-95){y}
    \end{picture}

	\includegraphics[scale=.8]{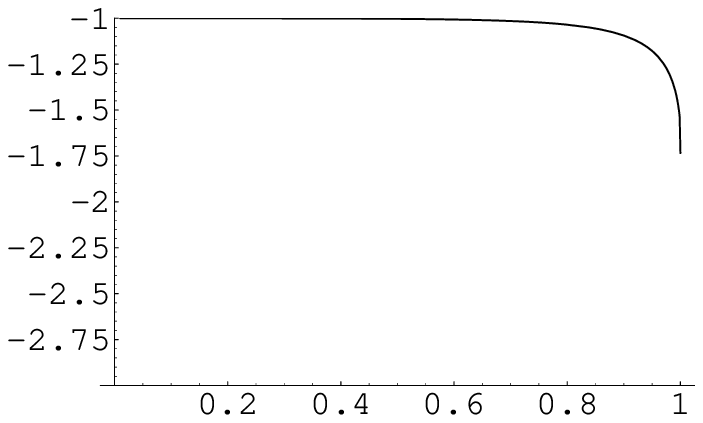}
	\caption{a = 10, t = -0.1}
    \label{bigcrnch4}

    \end{minipage}
\end{figure}

\begin{figure}
 \vspace{.5in}
    \begin{minipage} [b] {.5\linewidth}
\begin{picture} (0,0)
    	\put(10,120){$\frac{\Psi}{\mathrm{A}}$}
	\put(175,85){y}
    \end{picture}
	\includegraphics[scale = .8]{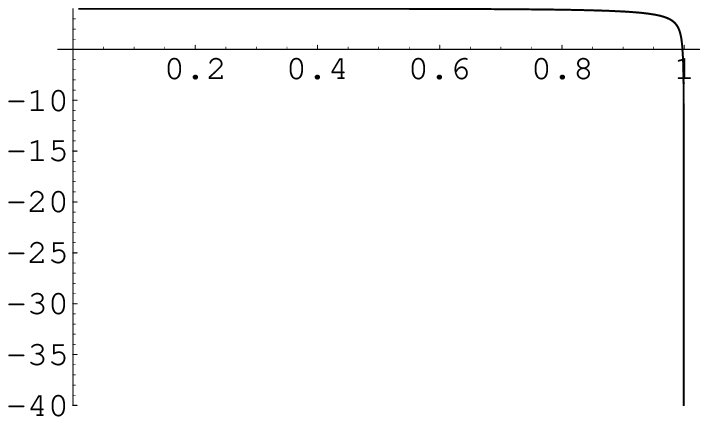}
    \caption{ a = 10, t = -$10^{-6}$}
    \label{bigcrnch5}

\end{minipage} %
    \begin{minipage} [b] {.5\linewidth}
\begin{picture} (0,0)
    	\put(160,-110){$\frac{\Psi}{\mathrm{A}}$}
	\put(0,0){y}
    \end{picture}

	\includegraphics[scale=.8]{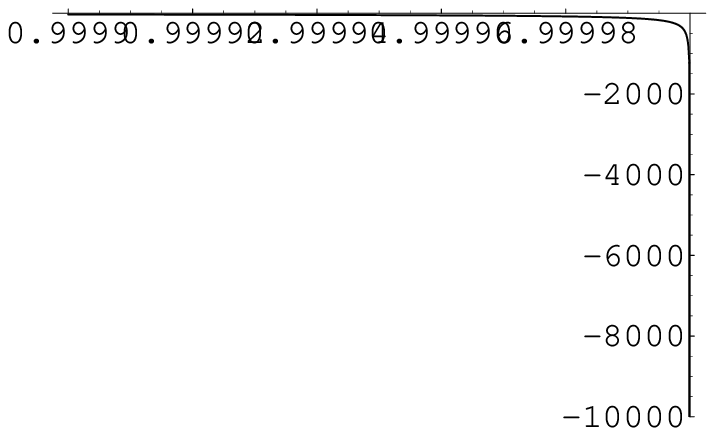}
	\caption{a = 10, t = -$10^{-6}$ (detail)}
    \label{bigcrnch6}

    \end{minipage}
\end{figure}

\indent On the other hand we are also interested in a solution whose 
boundary values goes to minus infinity at a finite time.  This 
corresponds in the boundary theory to sending the gauge coupling to 
zero.  Specifically we will consider
\begin{equation} \label{bigcrunch1}
\phi = -\frac{\mathrm{A}}{(e^{\mathrm{at}} - \mathrm{c})^{2}}
\end{equation}
which diverges as $e^{\mathrm{at}} \rightarrow \mathrm{c}$.  For 
$e^{\mathrm{at}} < c \, $ one gets
\begin{equation}\label{bigcrunch2}
\Psi = -\frac{\mathrm{A}}{\mathrm{c}^{2}} - \frac{\mathrm{A} 
\pi}{\mathrm{c}^{2}}\sum_{\mathrm{k} = 1}^{\infty} \Bigg( 
\frac{e^{\mathrm{at}}}{\mathrm{c}}\Bigg)^{\mathrm{k}}\frac{\frac{\mathrm{ak}}{2}(1+ \mathrm{k})(1+\frac{(\mathrm{ak})^2}{4})}{\sinh(\frac{\pi \mathrm{ak}}{2})}
\mathrm{F}(\frac{i\mathrm{ak}}{2},-\frac{i\mathrm{ak}}{2}, 2, \mathrm{y})
\end{equation}
The 
bulk solution gets arbitrarily large near the boundary as 
$e^{\mathrm{at}} \rightarrow \mathrm{c}$.  When the string coupling 
becomes of order $\frac{1}{\mathrm{N}}$ the string scale is 
comparable to the AdS scale and supergravity is no longer a reliable 
approximation.  Of course one generically encounters string scale 
curvature before running into a singularity.  There does not seem to be a simple analytic continuation in the 
mode sum formalism to t $ > \frac{\log{\mathrm{c}}}{\mathrm{a}}$.  
Note this $\phi$ also changes the boundary theory significantly; one is 
sending an effective coupling $g_{s}N$ from a large value to zero.  I should 
also note one can easily get a series of related solutions by 
taking derivatives with respect to c.  However, it is not entirely clear 
whether boundary values which diverges as $\frac{1}{(\mathrm{t} - 
\mathrm{t}_{0})^{\mathrm{n}}}$ for odd n make much sense for all times; 
after the divergence these functions send the coupling constant to 
infinity.  Figures~\ref{bigcrnch1}~-\ref{bigcrnch6} display the 
results for (\ref{bigcrunch2}) using the same approximation scheme
as before.  

\setcounter{equation} {0}
\section{Energy and Formation of Black Holes in $\mathbf{AdS}_{\mathrm{d} + 1}$}

\indent I now wish to discuss the energy propagating through the boundary. 
This quantity is nonzero once one allows time dependent boundary conditions.  
First, however, note that it is possible to get a good approximation for the field near the 
boundary from the equation of motion.  Let us define $\delta = \cos{\rho}^{2} = 1 - \mathrm{y} = (
\frac{1}{y} - 1)^{-\frac{1}{2}} = \frac{1}{1 + \mathrm{r}^{2}}$.  The 
normalizable modes which aren't determined by boundary conditions at 
infinity are of order $\delta^{\frac{\mathrm{d}}{2}}$.  For d $=$ 2
\begin{equation}\label{expansiond=2}
\Psi = \phi + a\delta\log{\delta} + b\delta + O(\delta^{2}\log{\delta}, \delta^{2})
\end{equation}
where $a = \frac{\ddot{\phi}}{4}$.  For d even and $16 \geq$ d $\geq 4$,
\begin{equation}
\Psi = \phi + \sum_{\mathrm{n} = 1}^{\frac{\mathrm{d}}{2} - 1} 
a_{\mathrm{n}}\delta^{\mathrm{n}} + 
a \delta^{\frac{\mathrm{d}}{2}} \log{\delta} + 
b\delta^{\frac{\mathrm{d}}{2}} + O(\delta^{\frac{\mathrm{d}}{2} + 1}\log{\delta}, \delta^{\frac{\mathrm{d}}{2} + 1})
\end{equation}
where
\begin{equation}
a_{1} =  \frac{\ddot{\phi}}{4(1-\frac{\mathrm{d}}{2})}, \quad a_{2} = 
\frac{a_{1} + \frac{\ddot{a}_{1}}{4}}{2(1-\frac{\mathrm{d}}{2})}, 
\quad a_{\mathrm{n} + 1} = \frac{\mathrm{n}^{2}a_{\mathrm{n}} + 
\frac{\ddot{a}_{\mathrm{n}}}{4}}{(\mathrm{n} + 
1)(1-\frac{\mathrm{d}}{2})} \mathrm \quad 2 \leq \mathrm{n} \leq 
\frac{\mathrm{d}}{2} - 2 
\nonumber\\
\end{equation}
and
\begin{equation}
 a = \frac{2}{\mathrm{d}}(\frac{\mathrm{d}^{2}}{4} - 
1)a_{\frac{\mathrm{d}}{2} - 1} + \frac{\ddot{a}_{\frac{\mathrm{d}}{2} - 
1}}{2\mathrm{d}} 
\end{equation}
For d = 1
\begin{equation}
\Psi = \phi + b\delta^{\frac{1}{2}} + O(\delta )
\end{equation}
while for odd d, d $\geq 3$
\begin{equation}
\Psi = \phi + \sum_{\mathrm{n} = 1}^{\frac{\mathrm{d} -1}{2}} 
c_{\mathrm{n}}\delta^{\mathrm{n}} + 
b \delta^{\frac{\mathrm{d}}{2}} + O(\delta^{\frac{\mathrm{d}+1}{2}})
\end{equation}
where
\begin{equation}
c_{1} = a_{1}, \quad c_{2} = 
\frac{c_{1} + \frac{\ddot{c}_{1}}{4}}{2(2-\frac{\mathrm{d}}{2})},
\quad c_{\mathrm{n} + 3} = \frac{ (\mathrm{n} + 
2)^{2}c_{\mathrm{n}+2} -\mathrm{n}(\mathrm{n} + 1)c_{\mathrm{n} +1} 
+\frac{\ddot{c}_{\mathrm{n} + 2}}{4}}{(\mathrm{n} + 
3)(1-\frac{\mathrm{d}}{2})} \quad \mathrm{n} \geq 0
\end{equation}
I should note that the coefficient $b$ contains three types of 
contributions: normalizable modes specified by a boundary conditions 
on a spacelike slice, sub-leading corrections from boundary conditions at spatial infinity, and 
the casuality properties of the solution (advanced, retarded, etc.).

\indent Since the metric for $AdS_{d+1}$ (3.\ref{AdSmetric} )
is independent of $\tau$, there is a timelike killing vector 
$\xi^{\mu} = \delta^{\mu}_{\tau}$.  The energy passing through the 
boundary of AdS is given by
\begin{equation} \label{Ebdy}
\mathrm{E}_{\mathrm{bdy}} = \int_{\mathrm{bdy}} 
\mathrm{T}_{\mu \nu} \, \xi^{\mu} n^{\nu}
\end{equation}
where $\mathrm{n}^{\nu}$ is a unit vector radially in and the energy 
momentum tensor 
$\mathrm{T}_{\mu \nu} = \nabla_{\mu} \Psi \nabla_{\nu} \Psi - 
\frac{1}{2} \mathrm{g}_{\mu \nu} \nabla_{\sigma} \Psi \nabla^{\sigma} 
\Psi $.  For spherically symmetric solutions the energy which goes 
through a cylinder of radius $1 -\delta$ between time $\mathrm{t}_{1}$ and  $\mathrm{t}_{2}$ 
is
\begin{equation} \label{Ebdysphere}
\mathrm{E}_{\mathrm{bdy}} = 2\Omega_{\mathrm{d} - 1} 
\frac{(1-\delta)^{\frac{\mathrm{d}}{2}}}{\delta^{\frac{\mathrm{d}}{2} - 1}} 
\int_{\mathrm{t}_{1}}^{\mathrm{t}_{2}} \mathrm{d}\tau 
\, \, \partial_{\delta}\Psi \, \partial_{\tau}\Psi
\end{equation}
where $\Omega_{\mathrm{d} - 1}$ is the area of the unit (d - 1)-sphere.
$\int_{\mathrm{t}_{1}}^{\mathrm{t}_{2}} \mathrm{d}\tau 
\, \, \partial_{\delta}\Psi \, \partial_{\tau}\Psi \Bigg\arrowvert_{\delta = 
0} $ is generically finite and non-zero.   In particular for the three 
examples noted above it is given by
\begin{equation}
\phi = \mathrm{A}e^{\mathrm{at}} \, \Rightarrow 
    \int_{\mathrm{t}_{1}}^{\mathrm{t}_{2}} \mathrm{d}\tau \,\,
\partial_{\delta}\Psi \, \partial_{\tau}\Psi \Bigg\arrowvert_{\delta = 
0} = \frac{(\mathrm{Aa})^{2}}{2}(e^{\mathrm{at}_{2}} - 
e^{\mathrm{at}_{1}}) \nonumber \\
\end{equation}

\begin{equation} \label{finiteperturbenergy}
\phi = \frac{\mathrm{A}}{e^{\mathrm{at}} + c} \, \Rightarrow 
\int_{-\mathrm{t}_{1} + 
\frac{\log{\mathrm{c}}}{\mathrm{a}}}^{\mathrm{t}_{2} + 
\frac{\log{\mathrm{c}}}{\mathrm{a}}} \mathrm{d}\tau \, \,
\partial_{\delta}\Psi \, \partial_{\tau}\Psi \Bigg\arrowvert_{\delta = 
0} = \frac{(\mathrm{Aa})^{2}}{128\mathrm{c}^{2}} \Bigg( 
\mathrm{Sech}^{4} (\frac{\mathrm{at}_{1}}{2})  - \mathrm{Sech}^{4} (\frac{\mathrm{at}_{2}}{2})) \Bigg), \nonumber\\
\end{equation}
\begin{equation}
\phi = \frac{\mathrm{A}}{(e^{\mathrm{at}} - c)^{2}} \, \Rightarrow 
\int_{\mathrm{t}_{1}}^{\mathrm{t}_{2}} \mathrm{d}\tau \, \,
\partial_{\delta}\Psi \, \partial_{\tau}\Psi \Bigg\arrowvert_{\delta = 
0} = \frac{(\mathrm{Aa})^{2}}{2} \Bigg(\frac{e^{\mathrm{2at}_{1}}}{(c 
- e^{\mathrm{at}_{1}})^{6}} - \frac{e^{\mathrm{2at}_{2}}}{(c - 
e^{\mathrm{at}_{2}})^{6}}\Bigg), \quad \quad \nonumber\\
\end{equation}
Then, due to the factor of 
$\frac{1}{\delta^{\frac{\mathrm{d}}{2} - 1}}$, generically we expect the 
energy going through the boundary over some small time interval to 
diverge for $AdS_{\mathrm{d}}, d \geq 4$ and to be finite for 
$AdS_{3}$.   Note, however, since the perpendicular area goes like $\approx 
\frac{1}{\delta^{\frac{\mathrm{d}}{2} - 1}}$ in all cases where 
$\Psi$ has bounded derivatives the energy per area going through the 
boundary is finite.  This would seem to confirm the proposition in 
the introduction that we get an infinite or at least a very, very large response 
in the bulk.  Of course, at some times this expression diverges to 
plus infinity and at others to minus infinity and we want to know 
the total energy which goes into the bulk.  Note, for example, (\ref{finiteperturbenergy}) is zero if we take $\mathrm{t}_{1} \, = 
\, \mathrm{t}_{2}$ (as well as trivially if $\mathrm{t}_{1} \, = 
\, -\mathrm{t}_{2}$) but is nonzero if $\vert \mathrm{t}_{1} \vert  \, \neq 
\, \vert \mathrm{t}_{2} \vert$ and leads to a divergence in 
(\ref{Ebdysphere}).  For generic $\phi \,$~using the expansions listed in the beginning of this section 
we can find the total energy over all time going through a cylinder of 
finite radius and take the limit as the radius goes to infinity.  The 
result is that the total energy going through the boundary over all 
time is finite and equal to
\begin{equation}
    \mathrm{E}_{\, \mathrm{total \, \, boundary}} =  
    \Omega_{\mathrm{d} - 1} \mathrm{d}\int_{-\infty}^{\infty} \mathrm{d}\tau 
    \, \,
    b \, \dot{\phi}
\end{equation}
if $\phi$ is bounded and $\phi^{(m)} \rightarrow 0$ for $m \geq 1$ 
as $\tau \rightarrow \pm \infty$.  Then provided $\phi$ approaches 
a constant at $\pm \infty$, not necessarily the same, finite 
changes in the gauge coupling result in a net finite energy change in 
the bulk.

\indent We can estimate the existence of black holes in the test field 
approximation by comparing the energy in a sphere of radius $\mathrm{r}_{H}$ to the energy of black 
hole of the same radius.  This estimate should be 
conservative---gravity makes things collapse.  Then a sufficient 
(although not necessary) condition for the formation of a black hole is
\begin{equation}
\mathrm{E}(\mathrm{r} \leq \mathrm{r}_{H}) \geq 
\mathrm{M}_{\mathrm{BH}}(\mathrm{r}_{H}) = \frac{(\mathrm{d} - 1)  
\Omega_{\mathrm{d} - 1} \mathrm{r}_{H}^{\mathrm{d} - 2} (1+\mathrm{r}_{H}^{2})}{16 \pi \mathrm{G}_{\mathrm{d} + 1}}
\end{equation}
The energy contained in a sphere of radius $\rho_{H}$ in the bulk is
\begin{equation}
\mathrm{E}_{\rho_{H}} = \int \mathrm{T}_{\mu \nu} \,
\xi^{\mu}\tilde{n}^{\nu} = \frac{\Omega_{\mathrm{d} - 1}}{2} 
\int_{0}^{\rho_{H}} \mathrm{d}\rho \, \tan^{\mathrm{d} - 1}(\rho) \, \, ( 
(\partial_{\tau} \Psi)^{2} + (\partial_{\rho} \Psi)^{2})
\end{equation}
where $\tilde{n}^{\nu}$ is a unit timelike vector.  Then setting 
$\mathrm{G}_{\mathrm{d} + 1} = 1$ (or absorbing it into the amplitude of 
$\Psi$) in terms of y =  $\sin^{2}(\rho)$, we get a 
black hole if
\begin{equation} \label{BHexist}
\int_{0}^{\mathrm{y}_{H}} \mathrm{dy} \, 
\frac{\mathrm{y}^{\frac{\mathrm{d}}{2} - 
1}}{(1-\mathrm{y})^{\frac{\mathrm{d}}{2}}} \, ( 
(\partial_{\tau} \Psi)^{2} + 
4\mathrm{y}(1-\mathrm{y})(\partial_{\mathrm{y}} \Psi)^{2}) \geq 
\frac{(\mathrm{d} - 1)}{4 \pi} 
\frac{\mathrm{y}_{H}^{\frac{\mathrm{d}}{2} - 1}}{(1- 
\mathrm{y}_{H})^{\frac{\mathrm{d}}{2}}}
\end{equation}
Note the divergence in the integrand as $\mathrm{y} \rightarrow 1$ is 
of the same type as the divergence in the right hand side and, since integrals 
make things less divergent, we do not get infinite black holes if 
$\Psi$ has bounded derivatives.  It is not hard to explicitly check 
this assertion for, e.g., $AdS_{5}$ and $AdS_{3}$.   This matches, of 
course, the fact that we only get a finite total amount of 
energy input into the bulk.

\begin{figure}
 \vspace{.1in}
    \begin{minipage} [b] {.5\linewidth}
	\includegraphics[scale = .8]{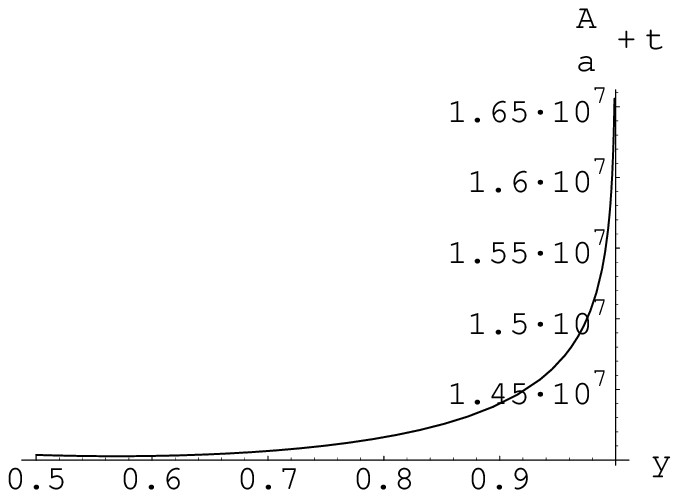}
    \caption{ a $= 10^{-6}$, y vrs. t + $\frac{\mathrm{A}}{\mathrm{a}}$}
    \label{ExpBH1}

\end{minipage} %
    \begin{minipage} [b] {.5\linewidth}
	\includegraphics[scale=.8]{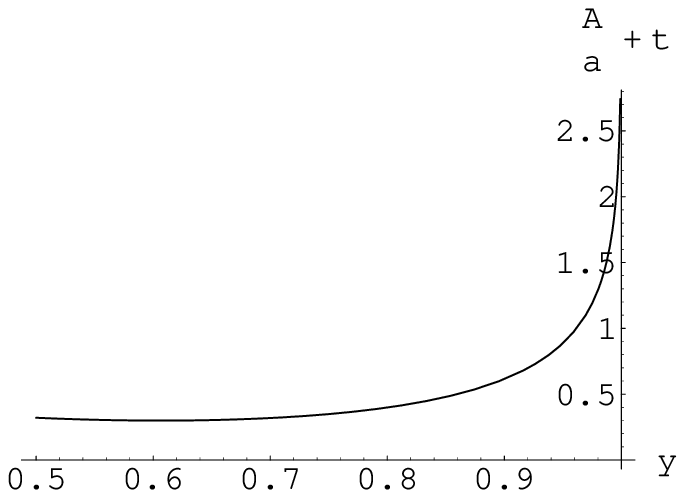}
	\caption{a $=$ 1, y vrs. t + $\frac{\mathrm{A}}{\mathrm{a}}$}
    \label{ExpBH2}

    \end{minipage}
\end{figure}

\begin{figure}
 \vspace{.1in}
    \begin{minipage} [b] {.5\linewidth}
	\includegraphics[scale = .8]{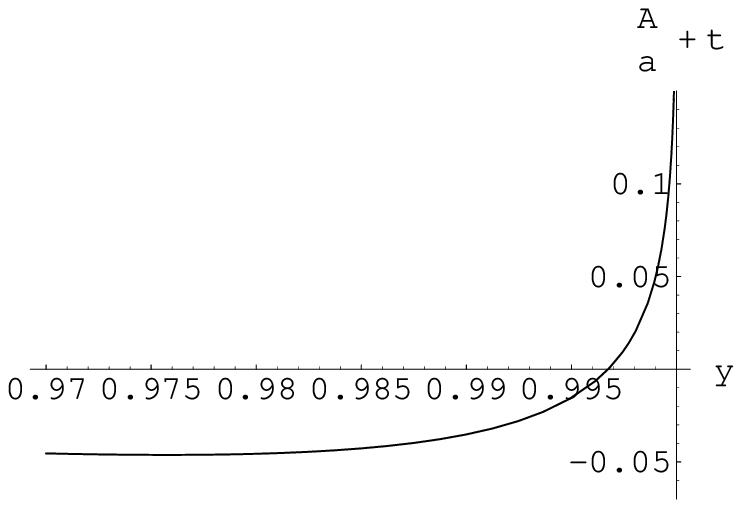}
    \caption{ a $= 10$, y vrs. t + $\frac{\mathrm{A}}{\mathrm{a}}$}
    \label{ExpBH3}

\end{minipage} %
    \begin{minipage} [b] {.5\linewidth}
	\includegraphics[scale=.55]{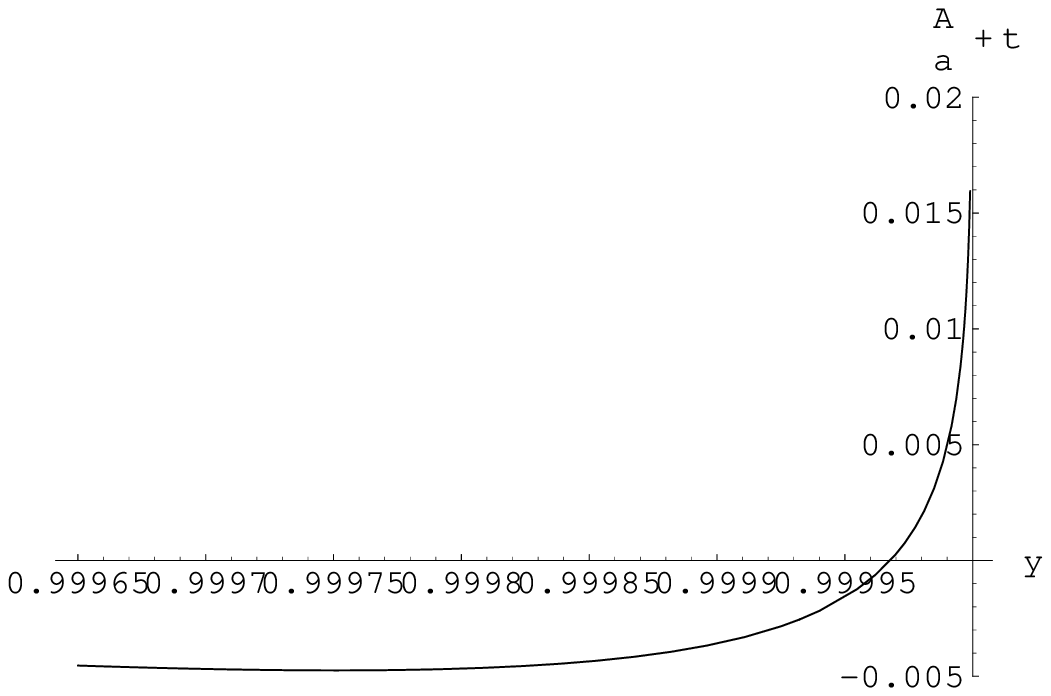}
	\caption{a $=$ 100, y vrs. t + $\frac{\mathrm{A}}{\mathrm{a}}$}
    \label{ExpBH4}

    \end{minipage}
\end{figure}

\indent Let us now examine the results for black hole formation for the 
previously discussed explicit examples.  
Figures~\ref{ExpBH1}~-~\ref{ExpBH4} show the time at which we first 
have enough energy to form a black hole of given radius for $\phi = 
Ae^{\mathrm{at}}$.  Note for the exponential solutions A can always be set to 1 by an appropriate choice of 
the time origin.  Even 
for slowly changing $\phi$~the field near the boundary changes more 
rapidly than that near the origin and we get at least medium size 
black holes.  As $a$ becomes large we quickly get very large black 
holes (compared to the AdS radius) 
with horizons which, as a function of time, asymptote to the 
boundary.    

\begin{figure}
 \hspace{-.40in}
    \begin{minipage} [b] {.5\linewidth}
\begin{picture} (0,0)
    	\put(23,0){$\Delta$}
	\put(190,-60){y}
    \end{picture}

	\includegraphics[scale = .8]{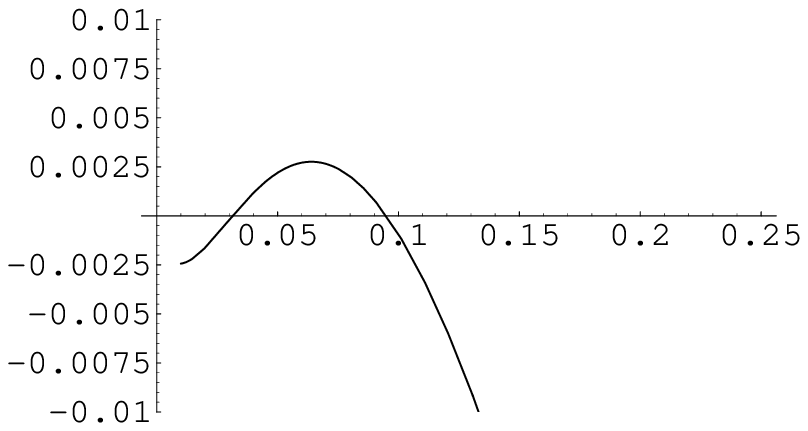}
    \caption{A $= 0.01$, t $= 1.4$}
    \label{tanhbh1}

\end{minipage} %
\hspace{.40in}
\begin{minipage} [b] {.5\linewidth}
    \begin{picture} (0,0)
    	\put(-15,-75){$\Delta$}
    \end{picture}

	\includegraphics[scale=.8]{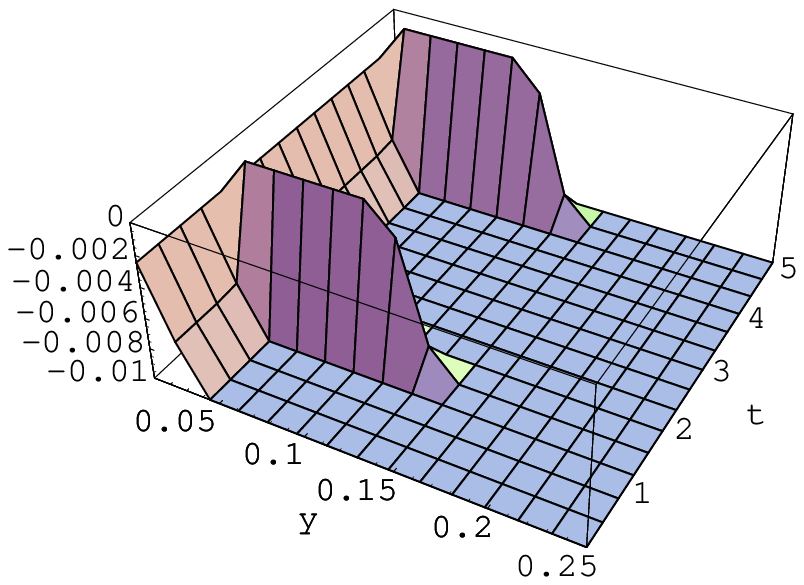}
	\caption{ A $= 0.01$}
    \label{tanhbh2}
    \end{minipage}
\end{figure}

\begin{figure}
    \begin{center}
 \begin{picture} (0,0)
    	\put(-108,-75){$\Delta$}
    \end{picture}

    \includegraphics[scale = .8]{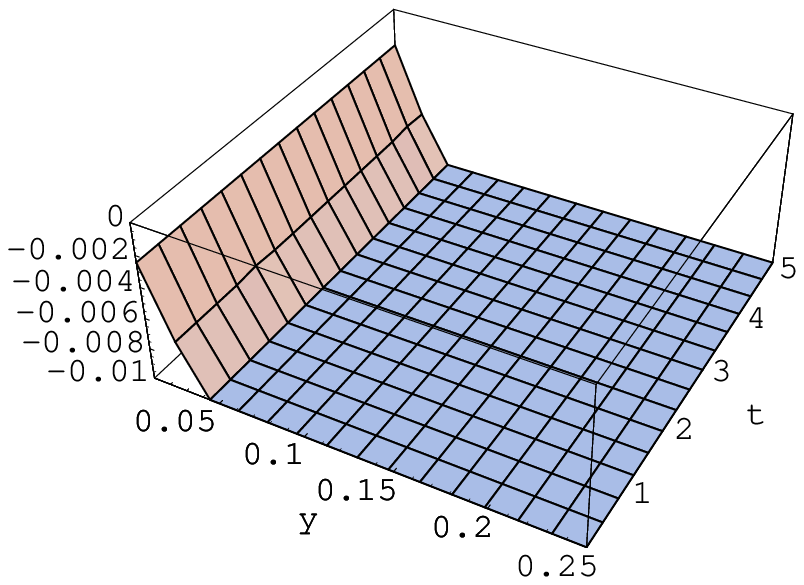}
    \caption{A $=0.001$}
    \label{tanhbh3}
\end{center}
\end{figure}

\indent For the finite perturbation discussed above we get finite size black 
holes.  By adjusting the parameters of $\phi$ we can adjust the size 
of the black hole.  Table \ref{table1} shows that by adjusting the 
overall amplitude we can determine the existence and size of a black 
hole at various times.  Define $\Delta$ as the left 
minus right side of 
(\ref{BHexist}); when $\Delta$ is positive there is enough energy to 
form a black hole of radius y at time t. 
Figures~\ref{tanhbh1}~-~\ref{tanhbh2} show we can get small black 
holes and Figure~\ref{tanhbh3} shows for a small enough amplitude as far as we can trust the 
approximation we don't ever get a black hole. 

\begin{table} \label{table1}
    \begin{center}
    \begin{tabular}{| l | c | r|}
    Amplitude & Time & BH size \\
    \hline
    A = 1  & t = 0.1 & $r_{BH}  \approx 5$ \\
    A = 0.25 & t = 0.25 & $r_{BH} \approx 2$ \\
    A = 0.2 & t = 0.25 & no BH \\
    A = 0.1 & t = 0.1 & no BH \\
    A = 0.1 & t = 0.25 & no BH \\
    A = 0.1 & t = 0.5 & no BH \\
     A = 0.1  & t = 0.65 & $r_{BH}  \approx 1.4$ \\
    \hline
    \end{tabular}
    \caption{Approximate horizon radii for various amplitudes A and 
    times t}
    \end{center}
    \end{table}

\indent On the other hand we wish to consider $\phi$ which has a pole 
of some order.  In particular, the divergent example mention above is dominated by
a 
second order pole near the divergence.    One finds in these 
diverging $\phi$ we get infinite black holes.  That is, given some small 
$\delta$ there is some time before the boundary value diverges at which the sphere of 
radius $1-\delta$ contains enough energy to form a black hole.  In 
particular, near the divergence $\phi \approx 
\frac{a}{(\mathrm{t} - \mathrm{t}_{0})^{b}}$.
Given \linebreak[3] $\delta \ll 1, \, \, \delta \ll \frac{(\mathrm{d} - 1)(\frac{\mathrm{d}}{2} - 1)}{4 \pi a^{2}b^{2} (1 - 2^{1 - 
\frac{\mathrm{d}}{2}})},$ and $
\delta \ll \Bigg( \frac{4 \pi a^{2}b^{2} (1 - 2^{1 - 
\frac{\mathrm{d}}{2}})}{(\mathrm{d} - 1)(\frac{\mathrm{d}}{2} - 1)} 
\Bigg)^{\frac{1}{b -1}}$ for $AdS_{\mathrm{d}} (\mathrm{d} \geq 4)$ the expansions in the beginning of the section can be 
trusted and taking the time  $\mathrm{t} = 
\mathrm{t}_{0} + \Bigg(\frac{4 \pi a^{2}b^{2} (1 - 2^{1 - 
\frac{\mathrm{d}}{2}}) \delta }{(\mathrm{d} - 1)(\frac{\mathrm{d}}{2} - 
1)} \Bigg)^{\frac{1}{2b}}$ upon expanding (\ref {BHexist}) for small 
$\delta$ one finds there is more than enough energy in a shell from 1 - 
2$\delta$ to 1 - $\delta$ to form a black hole of radius 1 - $\delta$.  For $AdS_{3}$ 
taking $\delta  \ll 1, \, \, \delta \ll \frac{1}{4 \pi a^{2}b^{2} 
\log{2}}, \, \,
\delta \log{\delta} \ll \Bigg( 4 \pi a^{2}b^{2} \log{2} \Bigg) ^{\frac{1}{b -1}}$
and time $\mathrm{t} = 
\mathrm{t}_{0} + \Bigg( 4 \pi a^{2}b^{2} \log{2} \Bigg) 
^{\frac{1}{2b}} $ 
leads to the desired conclusion.  Then these divergent 
solutions would seem to produce black holes that swallow up all 
of space and send the bulk into a big crunch.   The gauge theory is 
apparently
well behaved through the transition and so it should be possible 
to transmit information through at least this kind of collapse.  For 
the cases above after 
the divergence the dilaton returns to a finite 
value and at least this author thinks it likely one would have a 
sensible spacetime on the other side.   Definitive statements will 
require a better understanding of the AdS-FT dictionary.

\indent For small black holes obtaining analytic results seems nontrivial.   In particular, the taylor series one would write for small 
black holes for (\ref{BHexist}) doesn't converge quickly enough to be 
useful.  However, it is hopefully clear from the form of (\ref{BHexist}) 
and by the plotted examples that by choosing the parameters in $\phi$ 
we can determine the size of the black hole we make.  The spherically 
symmetric collapse of 
matter to form small black holes has been well studied numerically 
with a line of work started by Matthew Choptuik~\cite{Choptuik}.  These studies 
start with initial conditions on a spacelike slice but one would 
expect, especially for boundary conditions which change much faster 
than the AdS timescale, that one would quickly produce conditions very 
similiar to that work's initial conditions and hence reach a similiar conclusion.   In 
particular, the naked singularity found in that work should be 
resolved by string theory since we have an apparently perfectly well 
behaved dual quantum field theory. 

\section {Future Directions}

\indent As mentioned before the bulk calculations above are not 
expected to be qualitatively right because I have not included 
backreaction.  On this point I would like to solicit the attention of numerical 
relativists.  In terms of more basic theoretical issues, one needs a proper 
definition of asymptotically AdS and energy in the case where one has 
energy flowing through the boundary.  The usual references~\cite{Ashtekar}  on this 
issue assume the energy momentum tensor falls off at infinity faster 
than one finds in the cases I've discussed. 

\indent It remains an open 
question as to how often subtleties such as the ones described above 
prevent a sensible analytic continuation.  Virtually all of the work 
done on string theory in AdS and AdS/CFT in particular has been done in the Euclidean signature 
and it would be interesting to know how many more suprises await those 
examining the Lorentzian case.   Regardless, one has from the 
observations 
here a whole new 
AdS-FT dictionary to work out.  If this could be done one would almost 
certainly have many interesting things to say about matter collapsing 
to form black holes, singularity resolution, and possibly a
cosmological bounce.   On the other hand, finding even the standard 
AdS-CFT dictionary has proven to be a nontrivial task.  Although not 
the focus of this paper, using the bulk to study the field theory 
might also prove interesting; one has a way of at least numerically 
studying a strongly coupled, non-conformal, 
non-supersymmetric gauge theory.   One might also wonder whether one 
could find a scattering matrix formalism for the dual field 
theory in the case where one has a time dependent (although perhaps 
very slowly varying) gauge coupling.  Perturbing boundary conditions 
provides a rather direct link between the gauge and bulk theories and if 
we are clever and steadfast enough it may lead to some very 
interesting physics.       

\vspace{.75in}

\centerline{{\bf Acknowledgements}}
\vspace{.1in}
I would like to express my appreciation to Gary Horowitz suggesting 
this project and providing guidance throughout.  I'd also like to thank Joe Polchinski 
for several useful comments.

\end{document}